\documentclass[nospthms]{svjour3}    

\usepackage{url}
\newcommand{\urlBiBTeX}[1]{\url{#1}}

\usepackage{verbatim}
\usepackage{dsfont}
\usepackage{array}
\usepackage{cite}     
\usepackage{paralist}
\usepackage{amsfonts}
\usepackage{amssymb}
\usepackage{amsmath}
\usepackage{booktabs}
\usepackage{graphicx}
\usepackage[hyperref,amsmath,thmmarks]{ntheorem}

\def\Cfont#1{\textsf{#1}}

\def\sigA{\sigma^\text{\upshape a}}
\def\sigF{\sigma^\text{\upshape f}}

\def\sigAmain{\sigma^\text{\upshape act}}

\let\e \varepsilon

\def\IF{\ \Cfont{if}\ }
\def\IFF{\ \Cfont{iff}\ }
\def\after{\ \Cfont{after}\ }
\def\caused{\Cfont{caused}\ }
\def\causes{\ \Cfont{causes}\ }
\def\inertial{\Cfont{inertial}\ }

\def\default{\Cfont{default}\ }

\def\tval{\Cfont{t}}  
\def\fval{\Cfont{f}}

\def\drs{$D^{RS}$}

\def\cplus{$\mathit{C+}$}

\def\ccalc{\textsc{Ccalc}}

\def\sba{$\mathit{sub_1}$}
\def\sbb{$\mathit{sub_2}$}
\def\sbc{$\mathit{sub_3}$}
\def\sbd{$\mathit{sub_4}$}
\def\sbe{$\mathit{sub_5}$}
\def\sbf{$\mathit{sub_6}$}

\DeclareMathOperator{\val}{=}  

\newenvironment{mysplit}%
  {\arraycolsep 0pt \begin{array}{l}}%
  {\end{array}}

\newtheorem{Property}{Property}[section]

\theorembodyfont{\upshape}
\theoremsymbol{\ensuremath{\square}}
\newtheorem*{proof}{Proof}{\bf}{\upshape}

\begin{document}

\title{A Formal Specification of Dynamic Protocols for Open Agent Systems
}
\author{Alexander Artikis}
\institute{Institute of Informatics \& Telecommunications, \\ National Centre for Scientific Research ``Demokritos'', \\ Athens 15310, Greece \\ E-mail: \texttt{a.artikis@iit.demokritos.gr, a.artikis@acm.org} }

\date{ }

\maketitle 

\begin{abstract}
Multi-agent systems where the agents are developed by parties with competing interests, and where there is no access to an agent's internal state, are often classified as `open'. 
The member agents of such systems may inadvertently fail to, or even deliberately choose not to, conform to the system specification. Consequently, it is necessary to specify the normative relations that may exist between the agents, such as permission, obligation, and institutional power.
The specification of open agent systems of this sort is largely seen as a design-time activity. Moreover, there is no support for run-time specification modification. Due to environmental, social, or other conditions, however, it is often required to revise the specification during the system execution. To address this requirement, we present an infrastructure for `dynamic' specifications, that is, specifications that may be modified at run-time by the agents. The infrastructure consists of well-defined procedures for proposing a modification of the `rules of the game', as well as decision-making over and enactment of proposed modifications.
We evaluate proposals for rule modification by modelling a dynamic specification as a metric space, and by considering the effects of accepting a proposal on system utility. Furthermore, we constrain the enactment of proposals that do not meet the evaluation criteria.
We employ the action language \cplus\ to formalise dynamic specifications, and the `Causal Calculator' implementation of \cplus\ to execute the specifications.
We illustrate our infrastructure by presenting a dynamic specification of a resource-sharing protocol.
\end{abstract}


\section{Introduction}

A particular kind of Multi-Agent System (MAS) is one where the member agents are developed by different parties that have conflicting goals, and where there is no access to an agent's internal state. A key characteristic of this kind of MAS, due to the globally inconsistent goals of its members, is the high probability of non-conformance to the specifications that govern the members' interactions. A few examples of this type of MAS are electronic marketplaces, virtual organisations, and digital media rights management applications. MAS of this type are often classified as `open'.

Open MAS can be viewed as instances of \emph{normative systems} \cite{jones93}. A feature of this type of system is that actuality, what is the case, and ideality, what ought to be the case, do not necessarily coincide. Therefore, it is essential to specify what is permitted, prohibited, and obligatory, and perhaps other more complex normative relations that may exist between the agents. Among these relations, considerable emphasis has been placed on the representation of \emph{institutional power} \cite{jones96}. This is a standard feature of any normative system whereby designated agents, when acting in specified roles, are empowered by an institution to create specific relations or states of affairs. Consider, for example, the case in which an agent is empowered by an institution to award a contract and thereby create a bundle of normative relations between the contracting parties. 

Several approaches have been proposed in the literature for the specification of open MAS. The majority of these approaches offer `static' specifications, that is, there is no support for run-time specification modification. 
In some open MAS, however, environmental, social or other conditions may favour, or even require, specifications that are modifiable during the system execution. Consider, for instance, the case of a malfunction of a large number of sensors in a sensor network, or the case of manipulation of a voting procedure due to strategic voting, or when an organisation conducts its business in an inefficient manner. Therefore, we present in this paper an infrastructure for `dynamic' specifications, that is, specifications that are developed at design-time but may be modified at run-time by the members of a system. The presented infrastructure is an extension of our work on static specifications \cite{artikisToCL,artikisIGPL}, and is motivated by `dynamic argument systems' \cite{brewka01}. These are argument systems in which, at any point in the disputation, agents may start a meta level debate, that is, the rules of order become the current point of discussion, with the intention of altering these rules.

Our infrastructure for dynamic specifications allows agents to alter the specification of a protocol $P$ during the protocol execution. $P$ is considered an `object' protocol; at any point in time during the execution of the object protocol the participants may start a `meta' protocol in order to decide whether the object protocol specification should be modified. Moreover, the participants of the meta protocol may initiate a meta-meta protocol to decide whether to modify the specification of the meta protocol, or they may initiate a meta-meta-meta protocol to modify the specification of the meta-meta protocol, and so on.

Unlike existing approaches on dynamic specifications, we place emphasis on the procedure with which agents initiate a meta protocol. 
We distinguish between successful and unsuccessful attempts to initiate a meta protocol by identifying the conditions in which an agent has the institutional power to propose a specification change. We evaluate an agent's proposal for specification change by modelling a dynamic specification as a \emph{metric space} \cite{bryant85}, and by taking into consideration the effects of accepting a proposal on system utility. We constrain the enactment of proposals that do not meet the evaluation criteria. Furthermore, we formalise procedures for role-assignment in a meta level, that is, we specify which agents may participate in a meta protocol, and the roles they may occupy in the meta protocol.

We employ a resource-sharing protocol to illustrate our infrastructure for dynamic specifications: the object protocol concerns resource-sharing while the meta protocols are voting protocols. In other words, at any time during a resource-sharing procedure the agents may vote to change the rules that govern the management of resources. The resource-sharing protocol was chosen for the sake of providing a concrete example. In general, the object protocol may be any protocol for open MAS, such as a protocol for coordination or e-commerce. Similarly, a meta protocol can be any procedure for decision-making over specification modification (argumentation, negotiation, and so on).

We encode dynamic MAS specifications in executable action languages. In this paper we employ the action language \cplus\ \cite{lifschitz01}, a formalism with explicit transition system semantics. 
The \cplus\ language, when used with its associated software implementation, the `Causal Calculator' (\ccalc), supports a wide range of computational tasks of the kind that we wish to perform on MAS specifications.

The remainder of this paper is structured as follows. First, we present the action language \cplus. Second, we review a static specification of a resource-sharing protocol, and show how \ccalc\ can be used to prove properties of the static specification. Third, we present a dynamic specification of the resource-sharing protocol and an infrastructure for modifying the protocol specification during the protocol execution. We then show how \ccalc\ can be used to prove properties of the infrastructure for dynamic specifications, as well as support run-time activities by computing the normative relations current at each time. Finally, we summarise our work, discuss related research, and outline directions for further work.

\section{The \cplus\ Language}\label{sec:c+}

\cplus, as mentioned above, is an action language with an explicit transition system semantics. We describe here the version of \cplus\ presented in \cite{lifschitz01}. 

\subsection{Basic Definitions}

A \emph{multi-valued propositional signature} is a set $\sigma$ of symbols called \emph{constants}, and
for each constant $c \in \sigma$, a non-empty finite set $\mathit{dom}(c)$ of symbols, disjoint from $\sigma$, called the \emph{domain} of $c$. For simplicity, in this presentation we will assume that every domain contains at least two elements.

An \emph{atom} of signature $\sigma$ is an expression of the form $c \val u$ where $c \in \sigma$ and $u \in \mathit{dom}(c)$. A Boolean constant is one whose domain is the set of truth values $\{\tval, \fval \}$. When $c$ is a Boolean constant we often write $c$ for $c \val \tval$ and $\neg c$ for $c \val \fval$. A \emph{formula} $\varphi$ of signature $\sigma$ is any propositional combination of atoms of $\sigma$. An \emph{interpretation} $I$ of $\sigma$ is a function that maps every constant in $\sigma$ to an element of its domain. An interpretation $I$ \emph{satisfies} an atom $c \val u$ if $I(c)\val u$. The satisfaction relation is extended from atoms to formulas according to the standard truth tables for the propositional connectives.  A \emph{model} of a set $X$ of formulas of signature $\sigma$ is an interpretation of $\sigma$ that satisfies all formulas in $X$. If every model of a set $X$ of formulas satisfies a formula $\varphi$ then $X$ \emph{entails} $\varphi$, written $X \models \varphi$.

\subsection{Syntax}\label{sec:c+syntax}

The representation of an action domain in \cplus\ consists of \emph{fluent} constants and \emph{action}
constants.
\begin{itemize}
\item Fluent constants are symbols characterising a state. 
They are divided into two categories: simple fluent constants and statically determined fluent constants. Simple fluent constants are related to actions by \emph{dynamic laws}, that is, laws describing a transition $(s_i,\e_i,s_{i+1})$ from a state $s_i$ to its successor state $s_{i+1}$. Statically determined fluent constants are characterised by \emph{static laws}, that is, laws describing an individual state, relating them to other fluent constants. Static laws can also be used to express constraints between simple fluent constants. Static and dynamic laws are defined below.

\item Action constants are symbols characterising state transitions.
In a transition \linebreak $(s_i,\e_i,s_{i+1})$, the transition label $\e_i$, also called an `event', represents the actions performed concurrently by one or more agents or occurring in the environment. Transitions may be non-deterministic. Action constants are used to name actions, attributes of actions, or properties of transitions as a whole.
\end{itemize}
An \emph{action signature} $(\sigF,\sigA)$ is a non-empty set $\sigF$ of fluent constants and a non-empty set $\sigA$ of action constants. An \emph{action description} $D$ in \cplus\ is a non-empty set of \emph{causal laws} that define a transition system of a particular type. A causal law can be either a \emph{static law} or a \emph{dynamic law}. A static law is an expression
\begin{equation} \label{eq:static} \caused F \IF G \end{equation}
where $F$ and $G$ are formulas of fluent constants. In a
static law, constants in $F$ and $G$ are evaluated on the same
state. A dynamic law is an expression
\begin{equation} \label{eq:dynamic} \caused F \IF G \after  H \end{equation}
where $F$, $G$ and $H$ are formulas such that every constant
occurring in $F$ is a simple fluent constant, every constant
occurring in $G$ is a fluent constant, and $H$ is any combination of
fluent constants and action constants. In a
transition from state $s_i$ to state $s_{i+1}$,
constants in $F$ and in $G$ are evaluated on $s_{i+1}$,
fluent constants in $H$ are evaluated on $s_i$
and action constants in $H$ are evaluated on the transition
itself. $F$ is called the \emph{head} of the static law
(\ref{eq:static}) and the dynamic law (\ref{eq:dynamic}).

The full \cplus\ language also provides \emph{action dynamic laws}, which are expressions of the form
\begin{equation*} \caused \alpha \IF H \end{equation*}
where $\alpha$ is a formula containing action constants only and $H$ is a formula of action and fluent constants. We will not use action dynamic laws in this paper and so omit the details in the interests of brevity.

The \cplus\ language provides various abbreviations for common forms of causal law. For example, a dynamic law of the form
\begin{equation*} \caused F \IF \top \after H \wedge \alpha \end{equation*}
where $\alpha$ is a formula of action constants is often abbreviated as
\begin{equation*} \alpha \causes F \IF H \end{equation*}
In the case where $H$ is $\top$ the above is usually written as 
$\alpha \causes F$.

When presenting the resource-sharing protocol specification, we will often employ the $\Cfont{causes}$ abbreviation to express the effects of the agents' actions. We will also employ the \cplus\ abbreviation
\begin{equation*}
\default F
\end{equation*}
which is shorthand for the static law
\begin{equation*}
\caused F \IF F
\end{equation*}
expressing that $F$ holds in the absence of information to the contrary. 

When it aids readability, we will write
\begin{equation*} \caused F \IFF G \end{equation*}
as a shorthand for the pair of static laws
\begin{equation*} \caused F \IF G \end{equation*}
and
\begin{equation*} \default \neg F\end{equation*}

Finally, we will express the inertia of a fluent constant $c$ over time as:
\begin{equation*} \textsf{inertial } c \end{equation*}
This is an abbreviation for the \emph{set} of dynamic laws of the form (for all values $u \in \mathit{dom}(c)$):
\begin{equation*} \caused c \val u \textsf{ if } c \val u \textsf{ after } c \val u \end{equation*}

A \cplus\ action description is a non-empty set of causal laws. Of particular interest is the sub-class of  \emph{definite} action descriptions. A \cplus\ action description $D$ is \emph{definite} if:
\begin{itemize}
\item the head of every causal law of $D$ is an atom or $\bot$, and
\item no atom is the head of infinitely many causal laws of $D$.
\end{itemize}
The \cplus\ action description in this paper will be definite.


\subsection{Semantics}\label{sec:c+semantics}

It is not possible in the space available here to give a full account of the \cplus\ language and its semantics. We trust that the \cplus\ language, and especially its abbreviations, are sufficiently natural that readers can follow the presentation of the case study in later sections. Interested readers are referred to \cite{lifschitz01a,lifschitz01} for further technical details. For completeness, we summarise here the semantics of \emph{definite} action descriptions ignoring, as we are, the presence of action dynamic laws (and assuming that the domain of every constant contains at least two elements). We emphasise the transition system semantics, as in \cite{mjsCplus:techreport}.

Every action description $D$ of \cplus\ defines a labelled transition system, as follows:
\begin{itemize}
\item
States of the transition system are interpretations of the fluent constants
$\sigF$. It is convenient to identify a state $s$ with the set of fluent
atoms satisfied by $s$ (in other words, $s \models f\val v$ if and only if $f \val v \in
s$ for every fluent constant $f$).

\noindent Let $T_{\text{static}}(s)$ denote the heads of all static laws in $D$ whose conditions are satisfied by $s$:
\begin{equation*}
T_{\text{static}}(s) =_{\text{def}}
   \{F \mid \text{static law \eqref{eq:static} is in $D$}, s \models G \}
\end{equation*}
For a definite action description $D$, an interpretation $s$ of $\sigF$ is a \emph{state of the transition system defined by $D$}, or simply, a \emph{state of $D$}, when
\begin{equation*}
s = T_{\text{static}}(s) \cup \mathit{Simple}(s)
\end{equation*}
where $\mathit{Simple}(s)$ denotes the set of simple fluent atoms satisfied by $s$. (So $s - \mathit{Simple}(s)$ is the set of statically determined fluent atoms satisfied by $s$.)

\item Transition labels of the transition system defined by $D$ are the interpretations of the action constants $\sigA$.

\noindent A \emph{transition} is a triple $(s,\e,s')$ in which $s$ is the initial state, $s'$ is the resulting state, and $\e$ is the transition label. Since transition labels are interpretations of $\sigA$, it is
meaningful to say that a transition label $\e$ satisfies a formula $\alpha$ of $\sigA$: when $\e \models \alpha$ we sometimes say that the transition $(s,\e,s')$ is of type $\alpha$.

\item
Let $E(s,\e,s')$ denote the heads of all dynamic laws of $D$ whose conditions
are satisfied by the transition $(s,\e,s')$:
\begin{equation*}
   E(s,\e,s') =_\text{def}
       \{F \mid \text{dynamic law \eqref{eq:dynamic} is in $D$},
                         s' \models G, s \cup \e \models H \}
\end{equation*}
For a definite action description $D$, $(s,\e,s')$ is a \emph{transition of $D$}, or in full, a \emph{transition of the transition system defined by $D$}, when $s$ and $s'$ are interpretations (sets of atoms) of $\sigF$ and
$\e$ is an interpretation of $\sigA$ such that: 
\begin{itemize}
\item  $s = T_{\text{static}}(s) \cup \mathit{Simple}(s)$ \quad($s$ is a state of $D$)
\item  $s' = T_{\text{static}}(s') \cup E(s,\e, s')$
\end{itemize}

\end{itemize}

For any non-negative integer $m$, a \emph{path} or \emph{history of $D$} of length $m$ is a sequence
\begin{equation*} s_0 \, \e_0 \, s_1 \, \dots \, s_{m-1} \, \e_{m-1} \, s_m \end{equation*} 
where $(s_0, \e_0, s_1), \, \dots \, , (s_{m-1}, \e_{m-1}, s_m)$ are transitions of $D$.

\subsection{The Causal Calculator}\label{sec:causal-calculator}

The Causal Calculator\footnote{\urlBiBTeX{http://userweb.cs.utexas.edu/users/tag/cc/}} (\ccalc) is a software implementation developed by the Action Group of the University of Texas for representing  action and change in the \cplus\ language, and performing a range of computational tasks on the resulting formalisations. 
The functionality of \ccalc\ includes computation of `prediction' (temporal projection) and planning queries. Action descriptions in \cplus\ are translated by \ccalc\ first into the language of \emph{causal theories} \cite{lifschitz01} and then into propositional logic. The (ordinary, classical) models of the propositional theory correspond to paths in the transition system described by the original action description in \cplus. 
To compute an answer to a query, \ccalc\ invokes a satisfiability (SAT) solver to find models of the propositional theory which also satisfy the query.
A detailed account of \ccalc's operation and functionality may be found in \cite{lifschitz01}. 

In the following sections we present a \cplus\ action description, \drs, expressing a specification of a resource-sharing protocol. More precisely, first we present a static specification of a resource-sharing protocol, and then we present an infrastructure for dynamic specification of a resource-sharing protocol. 
Moreover, we show how we use \ccalc\ to execute a protocol specification.

\section{A Static Resource-Sharing Protocol}\label{sec:resource-sharing}

We present a specification of a resource-sharing or \emph{floor control} protocol in the style of \cite{artikisToCL}. 
In the field of Computer-Supported Co-operative Work the term `floor control' denotes a service guaranteeing that at any given moment only a designated set of users (subjects) may simultaneously work on the same objects (shared resources), thus creating a temporary exclusivity for access on such resources. We present a `chair-designated' Floor Control Protocol, that is, a distinguished participant is the arbiter over the usage of a specific resource. For simplicity we assume a single resource.

The protocol roles are summarised below:

\begin{itemize}
\item \emph{Floor Control Server} (\emph{FCS}), the role of the only participant physically manipulating the shared resource.
\item \emph{Subject} (\emph{S}), the role of designated participants requesting the floor from the chair, releasing the floor, and requesting from the FCS to manipulate the resource. 
\item \emph{Chair} (\emph{C}), the role of the participant assigning the floor for a particular time period to a subject, extending the time allocated for the floor, and revoking the floor from the subject holding it.
\end{itemize}

The floor can be either `granted', denoting that a subject has exclusive access to the resource (by the chair), or `free', denoting that no subject currently holds the floor. In both cases the floor may or may not be requested by a subject (for example, the floor may be granted to subject $\mathit{S'}$ and requested by subject $\mathit{S''}$ at the same time). 

\begin{figure}[t]
	\includegraphics[width=.8 \textwidth]{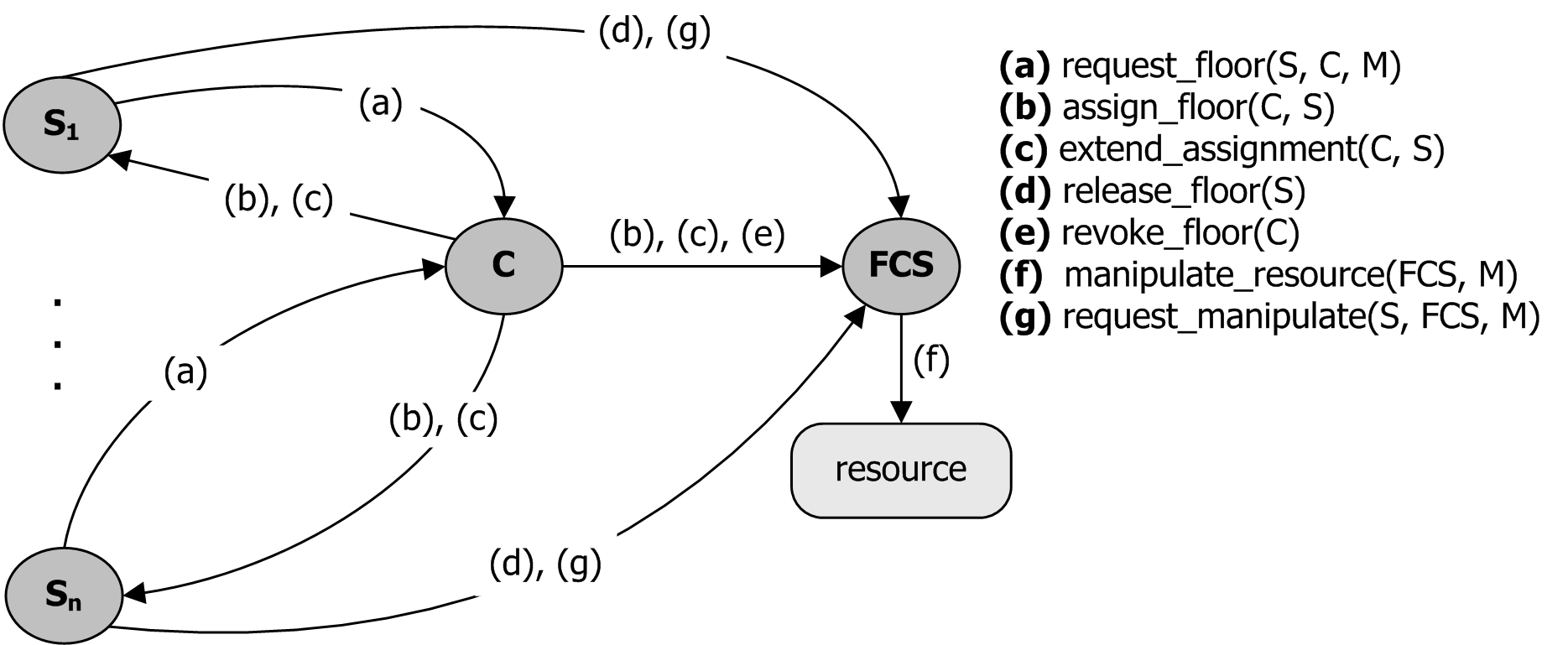}
	\caption{A Chaired Floor Control Protocol.}
	\label{fig:cfcp}
\end{figure}

Figure \ref{fig:cfcp} provides an informal description of the possible interactions between the agents occupying the protocol roles. More details about these interactions will be given presently.

Table \ref{tbl:c+signature} shows a subset of the action signature of \drs, that is, the \cplus\ the action description expressing the specification of the resource-sharing protocol. Variables start with an upper-case letter, and fluent and action constants start with a lower-case letter. The intended reading of the constants of the action signature will be explained below. 

\begin{table}[t]
\renewcommand{\arraystretch}{1} 
\caption{A Subset of the Action Signature of \drs\ (Part A).}\label{tbl:c+signature}
\begin{tabular}{p{1.1in}p{3.1in}}  \hline
\noalign{\smallskip}
Variable: & Domain:\\
$\mathit{M}$ & a set of resource manipulation types \\
$\mathit{Ag,\ S,\ S',\ C}$ & a set of agent ids \\ \\

\end{tabular}

\begin{tabular}{p{2.1in}p{2.1in}}  

Simple Fluent Constant: & Domain: \\
$\mathit{role\_of(Ag)}$ & $\mathit{ \{ subject, chair, fcs\} }$ \\
$\mathit{holder(S),\ sanctioned(Ag)}$ & Boolean \\ 
$\mathit{requested(S)}$ & a set of resource manipulation types \\
$\mathit{best\_candidate}$ &  a set of agent ids \\ 
$\mathit{c\_alloc(S)}$ &  $\mathds{Z}^+$ \\  \\

\end{tabular}

\begin{tabular}{p{3.2in}p{1in}}  

Statically Determined Fluent Constant (Boolean): & \\
$\mathit{powRequest(S, C),\ powAssign(C, S),\ powRequestMpt(S, FCS, M)}$ & \\ \\

\end{tabular}

\begin{tabular}{p{2.4in}p{1.8in}}  

Action Constant $\sigAmain$ (Boolean): \\ 
$\mathit{request\_floor(S, C, M),\ assign\_floor(C, S),\ request\_manipulate(S, FCS, M)}$ & \\ 
\hline
\end{tabular}
\end{table}

It has been argued \cite{jones96} that the specifications of protocols for open MAS should explicitly represent the concept of institutional power (or, for short, `power'), that is, the characteristic feature of institutions whereby designated agents, often when acting in specific roles, are empowered, by the institution, to create or modify facts of special significance in that institution --- \emph{institutional facts} --- usually by performing a specified kind of act. Searle \cite{searle65}, for example, has distinguished between \emph{brute facts} and institutional facts. Being in physical possession of an object is an example of a brute fact (it can be observed); being the owner of that object is an institutional fact. The resource-sharing protocol specification explicitly represents the concept of institutional power. Moreover, we follow the standard, long-established distinction between institutional power, physical capability, permission and obligation (see \cite{makinson86} for illustrations of this distinction). 

According to the resource-sharing protocol specification, all actions are physically possible at any time. In other examples the specification of physical capability could be different.

In this example, a subject $S$ is empowered to request the floor from the chair $C$ when $S$ has no pending requests:
\begin{equation} \label{eq:pow-req} 
\begin{mysplit}
\mathit{\caused powRequest(S, C) \IFF} \\
 	\qquad\mathit{role\_of(S) \val subject,} \\
       \qquad\mathit{role\_of(C) \val chair,} \\
       \qquad\mathit{requested(S) \val null}
\end{mysplit}
\end{equation}	
\cplus\ abbreviations, including \Cfont{iff}, were presented in Section \ref{sec:c+}.
The $\mathit{powRequest}$ fluent constant expresses the institutional power to request the floor, while the  $\mathit{role\_of}$ fluent constant expresses the role an agent occupies.  The $\mathit{requested}$ fluent constant records an agent's requests for the floor --- $\mathit{requested(S)\val null}$ denotes that $S$ has no requests for the floor. $\mathit{powRequest}$ is a statically determined fluent constant while $\mathit{role\_of}$ and $\mathit{requested}$ are simple fluent constants (see Table \ref{tbl:c+signature}). 

Each simple fluent constant of \drs\ is inertial, that is to say, its value persists by default from one state to the next. The constraint that a fluent constant $f$ is inertial is expressed in \cplus\ by means of the causal law abbreviation:
\begin{equation} \label{eq:c+inertia}
\inertial f
\end{equation}

Having specified the institutional power to request the floor, it is now possible to define the effects of this action: a request for the floor is eligible to be serviced if and only if it is issued by an agent with the institutional power to request the floor. Requests for the floor issued by agents without the necessary institutional power are ignored. Due to space limitations, we do not present here the \cplus\ laws expressing the effects of protocol actions (we show only one such law below).

We chose to specify that a subject is always permitted to exercise its power to request the floor. Moreover, a subject $S$ is permitted to request the floor even if $S$ is not empowered to do so. In the latter case a request for the floor will be ignored by the chair (since $S$ was not empowered to request the floor) but $S$ will not be \emph{sanctioned} since it was not forbidden to issue the request. In general, an agent is sanctioned when performing a forbidden action or not complying with an obligation. A few examples of sanctions will be shown presently (a more thorough treatment of sanctions may be found in \cite{artikisToCL, artikisIGPL}). Finally, a subject is never obliged to request the floor. 


The chair's power to assign the floor is defined as follows:
\begin{equation} \label{eq:pow-assign} 
\begin{mysplit}
\mathit{\caused powAssign(C, S) \IFF} \\
	\qquad\mathit{role\_of(C)\val chair,}\\
	\qquad\mathit{\forall S'\ \neg holder(S'),} \\
	\qquad\mathit{best\_candidate \val S}\\
\end{mysplit}
\end{equation}
The chair $C$ is empowered to assign the floor to $S$ if the floor is free, and $S$ is the best candidate for the floor. 
The simple fluent constant $\mathit{holder}$ expresses whether an agent has been allocated the floor. 
The simple fluent constant $\mathit{best\_candidate}$ denotes the best candidate for the floor. The definition of this constant is application-specific. For instance, the best candidate could be the one with the earliest request, that with the most `urgent' request (however `urgent' may be defined), and so on. 

The result of exercising the power to assign the floor to $S$ is that the floor becomes granted to $S$ for a specified time period. Sending an $\mathit{assign\_floor}$ message to an agent $S$  without the power to assign the floor to $S$ has no effect on the access rights of $S$. 

In this example, the conditions in which the chair is permitted to assign the floor are expressed as follows:
\begin{equation} \label{eq:per-assign} 
\begin{mysplit}
\mathit{\caused perAssign(C, S) \IFF}\\
       \qquad\mathit{role\_of(C)\val chair,}\\
	\qquad\mathit{\forall S'\ \neg holder(S'),} \\
       \qquad\mathit{best\_candidate \val S,} \\
	\qquad\mathit{c\_alloc(S)<3} \\
\end{mysplit}
\end{equation}	
The chair is permitted to assign the floor to $S$ if the floor is free, $S$ is the best candidate for the floor, and $S$ has not been allocated the floor the last 3 (or more) times. A simple fluent constant $\mathit{c\_alloc(S)}$ is incremented by 1 when $S$ is assigned the floor, and set to 0 when the floor is assigned to some other subject $S'$.

Note that, according to rules \eqref{eq:pow-assign} and \eqref{eq:per-assign}, the chair is not always permitted to exercise its power to assign the floor.

As mentioned above, an agent is subject to penalty when performing a forbidden action. In this case, a chair is subject to penalty when assigning the floor while being forbidden to do so. We record sanctions as follows:
\begin{equation} \label{eq:assign-sanction} 
\begin{mysplit}
\mathit{assign\_floor(C, S) \causes sanctioned(C) \IF}\\
	\qquad\mathit{role\_of(C)\val chair,}\\
       \qquad\mathit{\neg perAssign(C, S)}
\end{mysplit}
\end{equation}	
$\mathit{sanctioned}$ is a simple fluent constant. The actual penalty associated with the violation of a prohibition, or non-compliance with an obligation, may come in different flavours. We will show a type of penalty in a later section.

In this example, the conditions in which the chair is obliged to assign the floor are the same as the conditions in which the chair is permitted to assign the floor.

Similarly we specify the power, permission and obligation to perform the remaining protocol actions, and the effects of these actions. For instance, a subject's power to request a manipulation of the shared resource is defined as follows: 
\begin{equation} \label{eq:pow-req-man} 
\begin{mysplit}
\mathit{\caused powRequestMpt(S, FCS, M) \IFF}\\
	\qquad\mathit{role\_of(FCS)\val fcs,}\\
       \qquad\mathit{holder(S)}
\end{mysplit}
\end{equation}	
The $\mathit{powRequestMpt}$ fluent constant expresses the institutional power to request a resource manipulation.
A subject $S$ is empowered to request a resource manipulation of type $M$ from the floor control server $\mathit{FCS}$ if $S$ is the holder of the resource. 

The specification of the power, permission or obligation to request a resource manipulation, assign the floor, or perform some other protocol action, should include a deadline stating the time by which the action of requesting a resource manipulation, assigning the floor, etc, should be performed. Including deadlines in the formalisation lengthens the presentation and is omitted here for simplicity. Example formalisations of deadlines may be found in \cite{artikisToCL}.

\begin{table}[t]
\caption{A Resource-Sharing Protocol Specification.}\label{tbl:cfcp}
\renewcommand{\arraystretch}{1.2}
\setlength\tabcolsep{5pt}
\begin{tabular}{lccc}
\hline\noalign{\smallskip}
\multicolumn{1}{c}{\textbf{Action}} & \multicolumn{1}{c}{\textbf{Power}} & \multicolumn{1}{c}{\textbf{Permission}} & \multicolumn{1}{c}{\textbf{Obligation}} \\
\noalign{\smallskip}
\hline
\noalign{\smallskip}
$\mathit{request\_floor(S, C, M)}$ & $\mathit{requested(S)\val null}$ & $\top$ & $\bot$ \\[7pt]

$\mathit{assign\_floor(C, S)}$  & $\mathit{\forall S'\ \neg holder(S'),}$ & $\mathit{\forall S'\ \neg holder(S'),}$ & $\mathit{\forall S'\ \neg holder(S'),}$  \\[-2pt]
                                & $\mathit{best\_candidate = S}$  & $\mathit{best\_candidate = S,}$ & $\mathit{best\_candidate = S,}$ \\[-2pt]
				&  & $\mathit{c\_alloc(S)<3}$ & $\mathit{c\_alloc(S)<3}$ \\[7pt]
 													



$\mathit{request\_}$ & $\mathit{holder(S)}$ & $\mathit{holder(S)}$ & $\bot$ \\
$\mathit{manipulate(S, FCS, M)}$ &  &  &  \\

\hline
\end{tabular}
\end{table}

To summarise, Table \ref{tbl:cfcp} presents the conditions in which a protocol participant has the institutional power, permission and obligation to perform an action. To save space, in Table \ref{tbl:cfcp} we show only three protocol actions: $\mathit{request\_floor}$, $\mathit{assign\_floor}$ and $\mathit{request\_manipulate}$. Moreover, we do not display the $\mathit{role\_of}$ fluent constant and assume that $S$ denotes an agent occupying the role of subject, $C$ denotes an agent occupying the role of chair, and $\mathit{FCS}$ denotes an agent occupying the role of floor control server.

\section{Proving Properties of the Static Resource-Sharing Protocol}\label{sec:cfcp-properties}

The explicit transition system semantics of the \cplus\ language enables us to prove various properties of the presented specification, which is expressed by means of the action description \drs. Consider the following example.

\begin{Property}\label{prop:prop1} There is no protocol state in which the floor is free and the chair is not empowered to assign it to the best candidate. \end{Property}

\begin{proof} 
Assume a state $s$ of the transition system defined by \drs\ in which the best candidate for the floor is subject $S$, the floor is free, and the chair $C$ is not empowered to assign the floor to $S$. In other words, for any $S$, $C$, 
\begin{equation*}
\begin{mysplit}
\mathit{s\ \models\ best\_candidate\val S\ \wedge\ \forall S' \neg holder(S')\ \wedge\ role\_of(C)\val chair\ \wedge} \\
\qquad\quad\mathit{ \neg powAssign(C,S) }
\end{mysplit}
\end{equation*}

Since $s$ is a state of \drs, it is an interpretation of $\sigF$ such that 
$\mathit{s = T_{\text{static}}(s)\ \cup} \linebreak \mathit{Simple(s)}$, where
$
\mathit{T_{\text{static}}(s) =_{\text{def}} 
   \{F \mid \text{static law `$\caused F \IF G$' is in \drs},\ s \models G \} }
$
and $\mathit{Simple(s)}$ denotes the set of simple fluent atoms satisfied by $s$ (see Section \ref{sec:c+semantics}). From rule \eqref{eq:pow-assign}, and the fact that
\begin{equation*}
\mathit{s\ \models\ best\_candidate\val S\ \wedge\ \forall S' \neg holder(S')\ \wedge\ role\_of(C)\val chair }
\end{equation*}
we have that $\mathit{powAssign(C, S)} \in T_{\text{static}}(s)$.
According to our initial assumption, however, in $s$ the chair $C$ is not empowered to assign the floor to $S$, that is, \linebreak $\mathit{powAssign(C, S) \notin s}$, which implies that $s \neq T_{\text{static}}(s) \cup \mathit{Simple}(s)$. Therefore, $s$ is not a state of \drs. \end{proof} 

\ccalc\ provides an automated means for proving properties (such as Property \ref{prop:prop1}) of a protocol specification formalised in \cplus. We express the \cplus\ action description \drs\ in \ccalc's input language and then query \ccalc\ about \drs\ in order to prove properties of the protocol specification. (Details about the types of query that \ccalc\ computes, and \ccalc's input language, may be found in \cite{lifschitz01,lee01a,akman04a}.) Consider the following example.

\begin{Property}\label{prop:prop2}
A chair is always sanctioned when it performs a forbidden assignment of the floor.  \end{Property}

We instruct \ccalc\ to compute all states $s'$ such that 
\begin{itemize}
\item $\mathit{(s, \e, s')}$ is a transition of \drs, 
\item $\mathit{s\ \models\ \neg perAssign(C, S)\ \wedge\ role\_of(C)\val chair}$, and 
\item $\mathit{\e\ \models\ assign\_floor(C, S)}$.
\end{itemize}

For every state $s'$ computed by \ccalc\ we obtain
\begin{equation*} 
\mathit{s' \models sanctioned(C)}
\end{equation*}
This is due to rule \eqref{eq:assign-sanction}. 

Note that for some states $s'$ computed by \ccalc\ we obtain $\mathit{s' \models holder(S)}$, meaning that, in these cases, the chair $C$ was empowered, although forbidden, to assign the floor to subject $S$ (see, respectively, rules \eqref{eq:pow-assign} and \eqref{eq:per-assign} for the specification of the power and permission to assign the floor). \ccalc\ also computed states $s'$ such that $\mathit{s' \models \neg holder(S)}$, that is, in these cases, the chair was not empowered  to assign the floor to $S$. \vspace{9pt}

We may prove further properties of the resource-sharing protocol specification, in the manner shown above, such as that an agent is permitted to perform at least one action in every protocol state, an agent is never forbidden and obliged to perform an action, non-compliance with an obligation always leads to a sanction, and so on. Further examples of proving properties of protocol specifications formalised in \cplus\ will be presented in Section \ref{sec:dynamic-properties}.

\section{An Infrastructure for A Dynamic Resource-Sharing Protocol}\label{sec:dynamic-prot}

\begin{figure}[h]
	\includegraphics[width=.65 \textwidth]{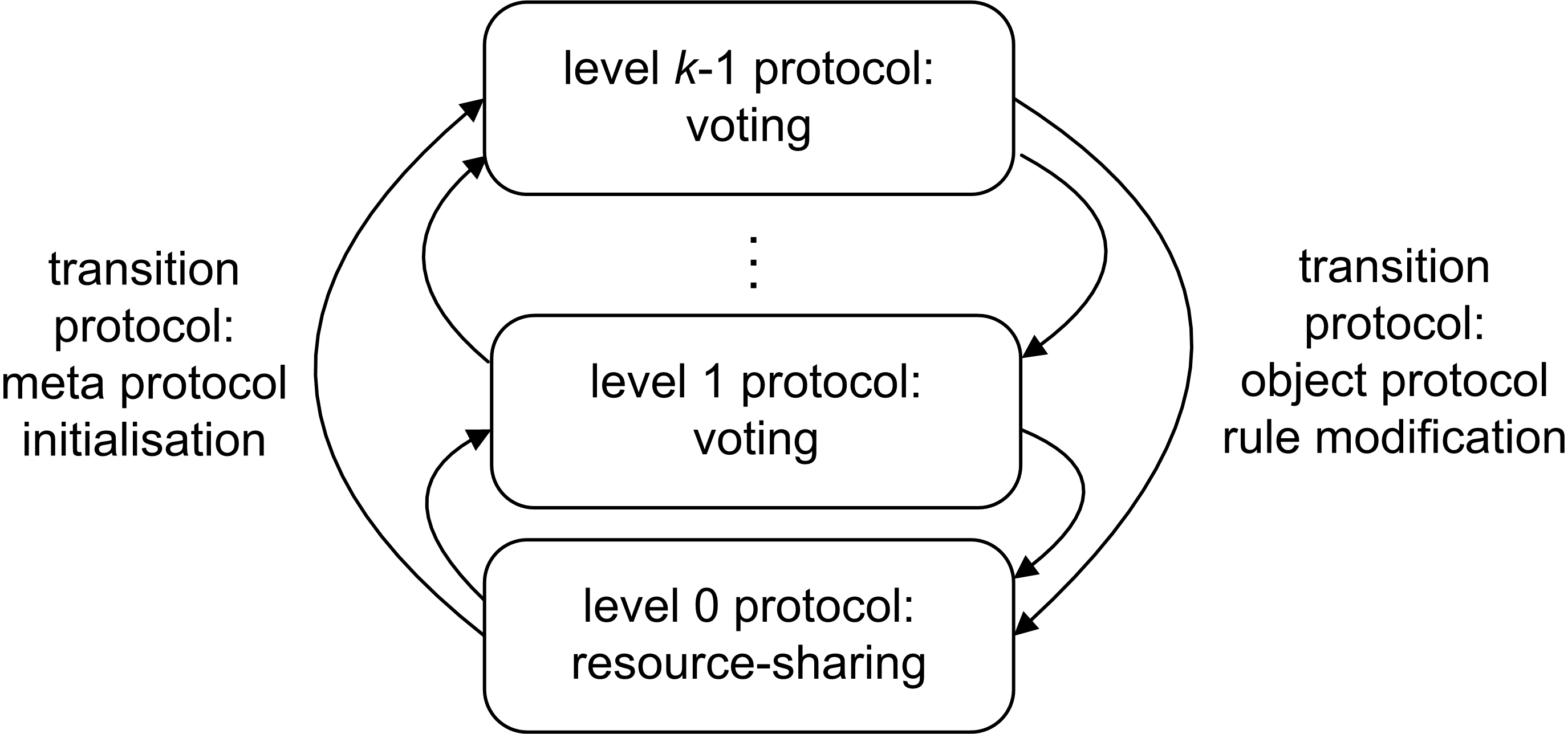}
	\caption{A $k$-level Infrastructure for Dynamic Specifications.}
	\label{fig:nlevel-protocol}
\end{figure}

Being motivated by Brewka \cite{brewka01}, we present an infrastructure that allows agents to modify (a subset of) the rules of a protocol at run-time. Regarding our running example, we consider the resource-sharing protocol as an `object' protocol; at any point in time during the execution of the object protocol the participants may start a `meta' protocol in order to potentially modify the object protocol rules --- for instance, replace an existing rule-set with a new one. The meta protocol may be any protocol for decision-making over rule modification. For the sake of presenting a concrete example, we chose a voting procedure as a meta protocol, that is, the meta protocol participants take a vote on a proposed modification of the object protocol rules. The participants of the meta protocol may initiate a meta-meta protocol to modify the rules of the meta protocol, or they may initiate a meta-meta-meta protocol to modify the rules of the meta-meta protocol, and so on. For simplicity, in this example \emph{all} meta protocols are voting procedures (in other systems each meta protocol may be a different decision-making procedure). In general, in a $k$-level infrastructure, level 0 corresponds to the main (resource-sharing, in this example) protocol, while a protocol of level $n$, $0{<}n{\leq}k{-}1$ (voting, in this example), is created, by the protocol participants of a level $m$, $0{\leq}m{<}n$, in order to decide whether to modify the protocol rules of level $n{-}1$. The infrastructure for dynamic (resource-sharing) specifications is displayed in Figure \ref{fig:nlevel-protocol}. 

\begin{table}[h]
\renewcommand{\arraystretch}{1} 
\caption{A Subset of the Action Signature of \drs\ (Part B). }\label{tbl:c+signature2}
\begin{tabular}{p{1.6in}p{2.8in}}  \hline
\noalign{\smallskip}
Variable: & Domain:\\
$\mathit{PL}$ & $ \mathds{Z}^+$  \\
$\mathit{SP, NSP, ASP, Motion}$ & a set of specification point ids \\
$\mathit{DoF\_ID}$ & a set of DoF ids \\ 
$\mathit{V}$ & $\mathit{ \{ for, against \} }$ \\
$\mathit{Outcome}$ & $\mathit{ \{ carried, not\_carried \} }$ \\ \\

\end{tabular}

\begin{tabular}{p{2.3in}p{2.1in}}  

Simple Fluent Constant: & Domain: \\
$\mathit{role\_of(Ag, PL)}$ & $\mathit{ \{ subject, chair, fcs\} }$ \\
$\mathit{dof(DoF\_ID, SP)}$ & a set of DoF values \\
$\mathit{actual\_sp(PL)}$ & a set of specification point ids \\
$\mathit{proposal(Ag, SP, PL),\ properties(SP, PL)}$ & Boolean \\ 
$\mathit{threshold\_d(PL),\ threshold\_eu(PL),}$ &  \\
$\mathit{eu(SP, PL)}$ & $\mathds{Z}^+$ \\  \\

\end{tabular}

\begin{tabular}{p{3.2in}p{1in}}  

Statically Determined Fluent Constant: & Domain:\\
$\mathit{powPropose(Ag, SP, PL),\ powSecond(Ag, SP, PL),}$ &  \\ 
$\mathit{perPropose(Ag, SP, PL),\ oblPropose(Ag, SP, PL),}$ &  \\
$\mathit{powDelcare(Ag, Motion, Outcome, PL)}$ & Boolean  \\
$\mathit{distance(SP, SP, PL)}$ & $\mathds{Z}^+$ \\ \\
\end{tabular}

\begin{tabular}{p{2.4in}p{1.8in}}  

Action Constant $\sigAmain$ (Boolean): & \\ 
$\mathit{propose(Ag, NSP, PL),\ second(Ag, NSP, PL),}$ & \\
$\mathit{vote(Ag, V, PL),\ declare(Ag, Motion, Outcome, PL)}$ & \\ 
\hline
\end{tabular}
\end{table}

Apart from object and meta protocols, the infrastructure for dynamic specifications includes `transition' protocols --- see Figure \ref{fig:nlevel-protocol} --- that is, procedures that express, among other things, the conditions in which an agent may successfully initiate a meta protocol (for instance, only the members of the board of directors may successfully initiate a meta protocol in some organisations), the roles that each meta protocol participant will occupy, and the ways in which an object protocol is modified as a result of the meta protocol interactions. The components of the infrastructure for dynamic specifications, level 0 protocol, level $n$ protocol ($n{>}0$), and transition protocol, are discussed in Sections \ref{sec:object-prot}--\ref{sec:spec-space}.

Table \ref{tbl:c+signature2} shows a set of fluent and action constants of the action signature of \drs\ that will be presented in the following sections. 
In order to distinguish between the protocol states of different protocol levels, we add a parameter, when necessary, in the representation of action and fluent constants, expressing the protocol level $\mathit{PL}$. For example, $\mathit{role\_of(Ag, PL)}$ expresses the role $\mathit{Ag}$ occupies in level $\mathit{PL}$. It is unnecessary to modify the syntax of action and fluent constants that are part of the specification of a single protocol level (such as the action constant $\mathit{request\_floor}$ that concerns only level 0).

\subsection{Level 0}\label{sec:object-prot}

For illustration purposes we chose a resource-sharing protocol --- the specification of which was presented in Section \ref{sec:resource-sharing} --- as a level 0 protocol. 
A protocol specification consists of the `core' rules that are always part of the specification, and the \emph{Degrees of Freedom} (\emph{DoF}), that is, the specification components that may be modified at run-time. (A DoF can be seen, for example, as Vreeswijk's `partial protocol specification' \cite{vreeswijk00}.)

A protocol specification with $l$ DoF creates an $l$-dimensional specification space where each dimension corresponds to a DoF. A point in the $l$-dimensional specification space, or \emph{specification point}, represents a complete protocol specification --- a \emph{specification instance} --- and is denoted by an $l$-tuple where each element of the tuple expresses a `value' of a DoF. Consider, for example, the resource sharing protocol with three DoF: the specification of the best candidate for the floor, the permission to assign the floor, and the permission to request a resource manipulation. The specification of these three protocol features may change at run-time --- for instance, the best candidate may be determined randomly, on a first-come, first-served basis, priority may be given to subjects requesting a particular manipulation type, or to subjects that have not been sanctioned (these are possible values of the best candidate DoF). Regarding the second DoF, it may be forbidden to assign the floor to subjects that have been allocated the floor the last 3, 6, or 9 times.
Finally, in this example, the holder may be permitted to request any type of resource manipulation from the Floor Control Server (FCS), or it may be permitted to request only the resource manipulation type expressed when it applied to the chair for the floor (these are possible values of the third DoF).
In the resource-sharing example with these three DoF, a specification point is, for instance:

\begin{center} $\mathit{(fcfs, 3, any\_type)}$ \end{center} 
According to the above specification point, the best candidate for the floor is determined on first-come, first-served basis ($\mathit{fcfs}$), the maximum number of permitted consecutive allocations of the floor to a subject is 3, while the holder is permitted to request any type of resource manipulation ($\mathit{any\_type}$).


In this example, the first DoF has 4 values, the second DoF has 3 values and the third DoF has 2 values. Consequently, the specification space contains 4$\times$3$\times$2$\val$24 specification points. In other examples we could have chosen different DoF (and/or DoF values), such as the specification of the permission and the obligation to request the floor, or perform any other protocol action. The classification of a specification component as a DoF is application-specific.

There are various reasons for which the agents may change at run-time the value of one or more DoF and thus the specification point. In the resource-sharing example, when the population of a system increases, the agents may decide to lower the limit of allowed consecutive allocations of the floor;  when the number of agents violating the laws increases, it may be decided to give priority to agents that have not been sanctioned when computing the best candidate for the floor; etc. Furthermore, an agent may attempt to change at run-time the value of a DoF in order to satisfy its own private goals. 

In any case, the value of one or more DoF, and thus the specification point, may change at run-time by means of a meta protocol. A discussion about meta protocols is presented in the following section.

We encode a protocol's specification points in \cplus\ as follows:
\begin{align} 
\label{eq:bc-configA} 
&\begin{mysplit}
\mathit{\caused dof(bc, sp_1) \val fcfs} \\
\mathit{\caused dof(per\_assign, sp_1) \val 9} \\
\mathit{\caused dof(per\_mpt, sp_1) \val any\_type}
\end{mysplit}
\\
\label{eq:bc-configB} 
&\begin{mysplit}
\mathit{\caused dof(bc, sp_2) \val rmt} \\
\mathit{\caused dof(per\_assign, sp_2) \val 9} \\
\mathit{\caused dof(per\_mpt, sp_2) \val expressed\_type}
\end{mysplit} 
\end{align} 
The simple fluent constant $\mathit{dof}$ records the value of a DoF --- $\mathit{dof(bc, sp_1) \val fcfs}$, for example, denotes that, when the protocol's specification point is $\mathit{sp_1}$, the best candidate ($\mathit{bc}$) for the floor is determined on a first-come, first-served basis.
$\mathit{per\_assign}$ denotes the DoF concerning the permission to assign the floor while $\mathit{per\_mpt}$ denotes the DoF concerning the permission to request a resource manipulation. 
$\mathit{rmt}$ is a value of the best candidate DoF, indicating that priority is given to subjects requesting a particular resource manipulation type.
The above formalisation shows two example specification points: $\mathit{sp_1\val(fcfs, 9, any\_type)}$ and $\mathit{sp_2\val(rmt, 9, expressed\_type)}$.

Each DoF value is defined by a set of rules --- consider the following example formalisation:
\begin{align} 
\label{eq:permpt-any} 
&\begin{mysplit}
\mathit{\caused perRequestMpt(S, FCS, M) \IF}\\
	\qquad\mathit{actual\_sp(0) \val ASP,}\\
	\qquad\mathit{dof(per\_mpt, ASP) \val any\_type,}\\
	\qquad\mathit{role\_of(FCS, 0)\val fcs,}\\
        \qquad\mathit{holder(S)}
\end{mysplit}
\\
\label{eq:permpt-req} 
&\begin{mysplit}
\mathit{\caused perRequestMpt(S, FCS, M) \IF}\\
	\qquad\mathit{actual\_sp(0) \val ASP,}\\
	\qquad\mathit{dof(per\_mpt, ASP) \val expressed\_type,}\\
	\qquad\mathit{role\_of(FCS, 0)\val fcs,}\\
        \qquad\mathit{holder(S),}\\
	\qquad\mathit{requested(S)\val M}
\end{mysplit} 
\end{align}
Rule \eqref{eq:permpt-any} defines the $\mathit{any\_type}$ value of the DoF concerning the permission to request a resource manipulation type, while rule \eqref{eq:permpt-req} defines the $\mathit{expressed\_type}$ value of this DoF. The $\mathit{perRequestMpt}$ fluent constant expresses the permission to request a resource manipulation of a particular type (for instance, storing files of a particular type on a shared storage device, or executing applications of a particular type on a shared processor).
According to rule \eqref{eq:permpt-any}, a subject $S$ is permitted to request any resource manipulation type $M$ if $S$ is the holder of the resource. 
According to rule \eqref{eq:permpt-req}, the holder $S$ of the resource is permitted to request only the manipulation type denoted when it applied to the chair for the floor (see the last condition of rule \eqref{eq:permpt-req} --- recall that the $\mathit{requested}$ fluent constant records the subjects' requests for the floor).
The simple fluent constant $\mathit{actual\_sp(PL)}$ records the actual specification point of protocol level $\mathit{PL}$. 
Rules \eqref{eq:permpt-any} and \eqref{eq:permpt-req} are replaceable in the sense that the participants of the resource-sharing protocol may active/deactivate one of them, at run-time, by changing the specification point in a way that the value of the DoF concerning the permission to request a resource manipulation type is modified. For example, moving from specification point $\mathit{sp_1}$ to $\mathit{sp_2}$ deactivates rule \eqref{eq:permpt-any} and activates rule \eqref{eq:permpt-req}. The ways in which a specification point may change at run-time are presented next.

\subsection{Level $n$}\label{sec:meta-prot}

To provide a concrete example, we chose a three-level infrastructure for dynamic specifications. Moreover, both levels 1 and 2 are voting procedures, such as that presented in \cite{pitt06cj}. 
A presentation of a voting procedure specification is beyond the scope of this paper.
%
Briefly, we assume a simple procedure including a set of voters casting their votes, `for' or `against' a particular motion, which would be in this example a proposed specification point change in level $n{-}1$, and a chair counting the votes and declaring the motion carried or not carried, based on the \emph{standing rules} of the voting procedure --- simple majority, two-thirds majority, etc.

Our infrastructure allows for the modification of the specification of all protocol levels apart from the top one. Consequently, we define DoF for all protocol levels apart from the top one. For the protocol of level 1 --- voting --- we chose to set as a DoF the standing rules of the voting procedure. In other words, a level 2 protocol may be initiated in order to decide whether level 1 voting should become, say, simple majority instead of two-thirds majority.
Note that the voting procedures of levels 1 and 2 may not always have the same set of rules. For example, at a particular time-point level 2 voting may require a simple majority whereas level 1 voting may require a two-thirds majority (as mentioned above, the standing rules of level 1 constitute a DoF and thus the specification of this part of the protocol may change at run-time).

\subsection{Transition Protocol}\label{sec:trans-prot}

In order to change the specification point of level $m$, $m{\geq}0$ (for example, to change the value of the best candidate DoF from $\mathit{fcfs}$ to $\mathit{rmt}$), that is, in order to start a protocol of level $\mathit{m{+}1}$, the participants of level $m$ need to follow a `transition' protocol --- see Figure \ref{fig:nlevel-protocol}. The infrastructure for dynamic specifications presented in this paper requires two types of transition protocol: one leading from the resource-sharing protocol to a voting one (level 0 to level 1 or 2), and one leading from one voting protocol to the other (level 1 to level 2). We will only present the first type of transition protocol; the latter type of protocol may be specified in a similar manner. An example transition protocol leading from resource-sharing to voting can be briefly described as follows. A subject $\mathit{S}$ of the resource-sharing protocol  proposes that the specification point of this level (or of level 1) changes. 
If $\mathit{S}$ is empowered to make such a proposal, then the modification is directly accepted, without the execution of a meta protocol, provided that another subject exercises its power to second the proposal, and no other subject exercises its power to object to the proposal. If $\mathit{S}$ is not empowered to make the proposal, or if the proposal is not seconded, then it is ignored. If the proposal is seconded, and there is an objection, then an argumentation procedure commences, the topic of which is the proposed specification point change, the proponent of the topic is $\mathit{S}$, the subject that made the proposal, and the opponent is the subject that objected to the proposal. The argumentation procedure is followed by a meta protocol (level 1 or 2) in which a vote is taken on the proposed specification point change.

In this example transition protocol we have specified the power to propose a specification point change as follows:
\begin{equation} \label{eq:propose-replace} 
\begin{mysplit}
\mathit{\caused powPropose(Ag, NSP, PL) \IF}\\
       \qquad\mathit{role\_of(Ag, 0) \val subject,}\\
       \qquad\mathit{actual\_sp(PL) \val ASP,\ ASP \neq NSP,}\\	    
       \qquad\mathit{protocol(PL{+}1) \val idle,}\\         
       \qquad\mathit{properties( NSP, PL )}
\end{mysplit}
\end{equation}
An agent $\mathit{Ag}$ is empowered to propose that the specification point of protocol level $\mathit{PL}$ becomes $\mathit{NSP}$ if the following conditions are satisfied. First, $\mathit{Ag}$ occupies the role of subject in level 0. In this example, the chair of the resource-sharing protocol (level 0) is not empowered to propose a change of the specification point. Second, the actual specification point, $\mathit{ASP}$, is different from $\mathit{NSP}$. Third, there is no protocol taking place in level $\mathit{PL{+}1}$. A $\mathit{protocol(PL)}$ simple fluent constant records whether a protocol of level $\mathit{PL}$ is idle or executing. A protocol for changing the specification point of level $\mathit{PL}$, that is, a protocol of level $\mathit{PL{+}1}$, may commence only if there is no other protocol of level $\mathit{PL{+}1}$ taking place.

Fourth, the specification instance corresponding to specification point $\mathit{NSP}$ of level $\mathit{PL}$ satisfies a set of properties --- $\mathit{properties}$ is a simple fluent constant. 
The properties that a specification instance should satisfy are application-specific. We may require, for example, that an agent in never forbidden and obliged to perform the same action, a floor control mechanism is `safe' and `fair' \cite{dommel97}, and so on.
In this example, we have chosen to specify that an agent is not empowered to propose the adoption of a specification point of the form $\mathit{(rmt, *, any\_type)}$ ($*$ denotes any value of the second DoF),  since these points correspond to specification instances that are, in some sense, `inconsistent': 
priority for access to the resource is given to subjects with an expressed request of resource manipulation $M$ (the value of the best candidate DoF is $\mathit{rmt}$), while it is permitted to actually request from FCS a different type of manipulation $M'$ (the value of the  DoF concerning the permission for actual resource manipulation is $\mathit{any\_type}$).

\ccalc\ is an automated reasoning tool allowing for proving protocol properties --- recall that in Section \ref{sec:cfcp-properties} we proved properties of a specification instance of the resource-sharing protocol by means of \ccalc's query computation. 
Assuming that a protocol's specification points are known before the commencement of the protocol execution, we may determine at design-time, with the use of \ccalc, whether the specification instance corresponding to each specification point of a protocol level satisfies a set of desirable properties. 
Accordingly, we may set the value of the $\mathit{properties}$ fluent constant (which is part of the transition protocol specification), thus avoiding, at run-time, the (time-consuming) task of proving protocol properties.

In other examples the specification of the power to propose a specification point change could have different, or additional conditions further constraining how a protocol specification may change at run-time. For instance, it may be required that, for any protocol level, the specification point does not change `too often', or there may be an upper limit on the number of specification point changes (proposed by an agent). 
In Section \ref{sec:spec-space} we present a way of further constraining the process of specification point change. Other ways of achieving that, such as the ones described above, could have been formalised.

A proposal for specification point change needs to be seconded by another subject having the institutional power to second the proposal in order to be directly enacted, or initiate the argumentation procedure leading to voting. We chose to specify the power to second a proposal for specification point change as follows:
\begin{equation} \label{eq:second} 
\begin{mysplit}
\mathit{\caused powSecond(Ag, NSP, PL)  \IF} \\
	\qquad \mathit{role\_of(Ag, 0) \val subject, } \\
	\qquad \mathit{proposal(Ag', NSP, PL),}\\
	\qquad\mathit{\ Ag\neq Ag'} 
\end{mysplit}
\end{equation}	
$\mathit{Ag}$ is empowered to second any proposal for specification point change made by some other $\mathit{Ag'}$. The $\mathit{proposal}$ simple fluent constant records proposals made by empowered agents (subjects, in this example). 

We have specified that any subject is empowered to object to a proposal for specification point change.

Exercising the power to object to a proposal for specification point change initiates an argumentation procedure, the topic of which is the proposed change. To save space we do not present here a specification of an argumentation procedure; see \cite{artikis06AIJ} for an example formalisation of such a procedure.

The completion of the argumentation taking place in the context of a transition protocol initiates a meta protocol. The latter protocol is a voting procedure concerning a proposed specification point change. The agents participating in this procedure and the roles they occupy are determined by the transition protocol that results in the voting procedure. We chose to specify that the chair of the resource-sharing protocol becomes the chair of the voting procedure. Furthermore, the agents occupying the role of voter, thus having the power to vote, are the subjects that have not been sanctioned for exhibiting `anti-social' behaviour, that is, performing forbidden actions or not complying with obligations:
\begin{equation} \label{eq:meta-role-ass} 
\begin{mysplit}
\mathit{\caused role\_of(Ag, PL)\val voter \IF}\\
	\qquad\mathit{role\_of(Ag, 0)\val subject,}\\
	\qquad\mathit{protocol(PL)\val executing,\ PL>0,}\\
	\qquad\mathit{\neg sanctioned(Ag)}
\end{mysplit}
\end{equation}
The value of a $\mathit{protocol(PL)}$ constant becomes $\mathit{executing}$, in the case where $\mathit{PL}{>}0$, at the end of the argumentation of the transition protocol that led to level $\mathit{PL}$.
We chose not to relativise the constant recording sanctions to a protocol level. Therefore, the simple fluent constant $\mathit{sanctioned}$ records `anti-social' behaviour exhibited at any protocol level. Depending on the employed treatment of sanctions, `anti-social' behaviour may be permanently recorded, thus permanently depriving a subject of participating in a meta level, or it may be temporarily recorded, thus enabling subjects to participate in a meta level after a specified period has elapsed from the performance of forbidden actions or non-compliance with obligations.

Clearly, meta level role-assignment may be specified in other ways. For example, it may be the case that agents that recently joined the system are not admitted in the meta level, or that priority is given to subjects that have had the least time accessing the shared resources, etc. In general, meta level role-assignment can be as complex as required by the application under consideration.

The fact that an agent may successfully start a protocol of level $\mathit{m{+}1}$ by proposing a change of the specification point of level $m$, does not necessarily imply that the specification point of level $m$ will be changed. It is only if the motion of level $m{+}1$ --- which is the proposed specification point for level $\mathit{m}$ --- is carried, that the specification point of level $m$ will be changed. Consider the following rule expressing the outcome of a voting procedure of level $m{+}1$, that takes place in order to change the specification point in level $m$ to $\mathit{NSP}$:
\begin{align} 
\label{eq:declare-replace} 
\begin{mysplit}
\mathit{declare(VC, Motion, carried, PL{+}1) \causes actual\_sp(PL)\val NSP \IF}\\
	\qquad\mathit{powDeclare(VC, Motion, carried, PL{+}1),} \\
	\qquad\mathit{Motion\val NSP}
\end{mysplit}
\end{align}
%
Exercising the power to declare the motion carried in level $\mathit{PL{+}1}$ results in changing the specification point in level $\mathit{PL}$ to $\mathit{NSP}$, provided that the motion concerned the adoption of $\mathit{NSP}$. If the chair of the voting procedure $\mathit{VC}$ did not declare the motion carried, or was not empowered to make the declaration, then the specification point would not have changed in level $PL$. To save space we do not present here the specification of the power to make a declaration.

Exercising the power to declare the motion carried in a meta level $m$ may have additional effects.
For example, we may explicitly specify whether or not deactivating a rule, by changing the specification point of protocol level $m{-}1$, results in removing the effects of the rule that were produced prior to the rule deactivation. 
A discussion about specification change operations is presented in the last section of the paper.

\subsection{Constraining the Process of Run-time Specification Modification}\label{sec:spec-space}

As already mentioned, all protocol levels apart from the top one have DoF and, therefore, their specification may be modified at run-time. In this section we present ways of evaluating a proposal for specification point change. Moreover, we present ways of constraining the enactment of proposals that do not meet the evaluation criteria. 


One way of evaluating a proposal for specification point change is by modelling a dynamic protocol specification as a \emph{metric space} \cite{bryant85}. More precisely, we compute the `distance' between the proposed specification point and the actual point.
We constrain the process of specification point change by permitting proposals only if the proposed point is not too `far' from/`different' to the actual point. The motivation for formalising such a constraint is to favour gradual changes of a system specification. 
In what follows we describe how we may compute the distance between two specification points, and the conditions in which a proposal for specification point change is considered permitted.

We may follow two steps to compute the distance between two specification points $sp$ and $sp'$, each represented as an $l$-tuple, where each element of the tuple expresses a DoF value. 
First, we may transform $sp$ and $sp'$ to $l$-tuples of non-negative integers $qsp$ and $qsp'$ respectively.
To achieve that we can define an application-specific function $v$, that `ranks' each DoF value, that is, associates every DoF value with a non-negative integer.
The $i$-th element of $qsp$, $qsp_i$, has the value of $v(sp_i)$, where $sp_i$ is the $i$-th element of $sp$ (respectively, the $i$-th element of $qsp'$, $qsp'_i$, has the value of of $v(sp'_i)$, where $sp'_i$ is the $i$-th element of $sp'$). 
Second, we may employ a \emph{metric} (or \emph{distance function}), such as the Euclidean metric, to compute the distance between $qsp$ and $qsp'$ (the choice of a metric is application-specific --- see \cite{bryant85} for a list of metrics). Depending on the employed metric, we may add weights on the DoF --- for instance, we may require that the computation of the distance between $qsp$ and $qsp'$ is primarily based on the best candidate DoF rather than the other two DoF. The distance between $sp$ and $sp'$ is the distance between $qsp$ and $qsp'$.

Alternatively, we may compute the distance between two specification points by defining an application-specific metric that does not necessarily rely on a quantification DoF values.

We may constrain run-time specification point change as follows:
\begin{align} 
\label{eq:per-propose-metric} 
&\begin{mysplit}
\mathit{\caused perPropose(Ag, NSP, PL) \IFF}\\
       \qquad\mathit{powPropose(Ag, NSP, PL),}\\
       \qquad\mathit{actual\_sp(PL)\val ASP,}\\
	\qquad\mathit{distance( NSP, ASP, PL ) \leq  threshold\_d(PL) }
\end{mysplit}
\end{align}	
$\mathit{perPropose(Ag, NSP, PL)}$ is a statically determined fluent constant denoting whether agent $\mathit{Ag}$ is permitted to propose that the specification point of level $\mathit{PL}$ becomes $\mathit{NSP}$. 
$\mathit{distance}$ is a statically determined fluent constant computing the distance between any two specification points of a protocol level. The definition of $\mathit{distance}$ includes a \cplus\ formalisation of a chosen metric.
$\mathit{threshold\_d(PL)}$ is a simple fluent constant recording the maximum distance that a proposed specification point should have from the actual point in level $\mathit{PL}$.
Different protocol levels may have different $\mathit{threshold\_d}$ values and different metrics.
According to rule \eqref{eq:per-propose-metric}, an agent $\mathit{Ag}$ is permitted to propose that the specification point of level $\mathit{PL}$ becomes $\mathit{NSP}$, if and only if $\mathit{Ag}$ is empowered to make this proposal, and the distance between $\mathit{NSP}$ and the actual specification point of level $\mathit{PL}$, $\mathit{ASP}$, is less or equal to a specified threshold.
%

\begin{figure}[t]
  \begin{center}
	\includegraphics[width=\textwidth]{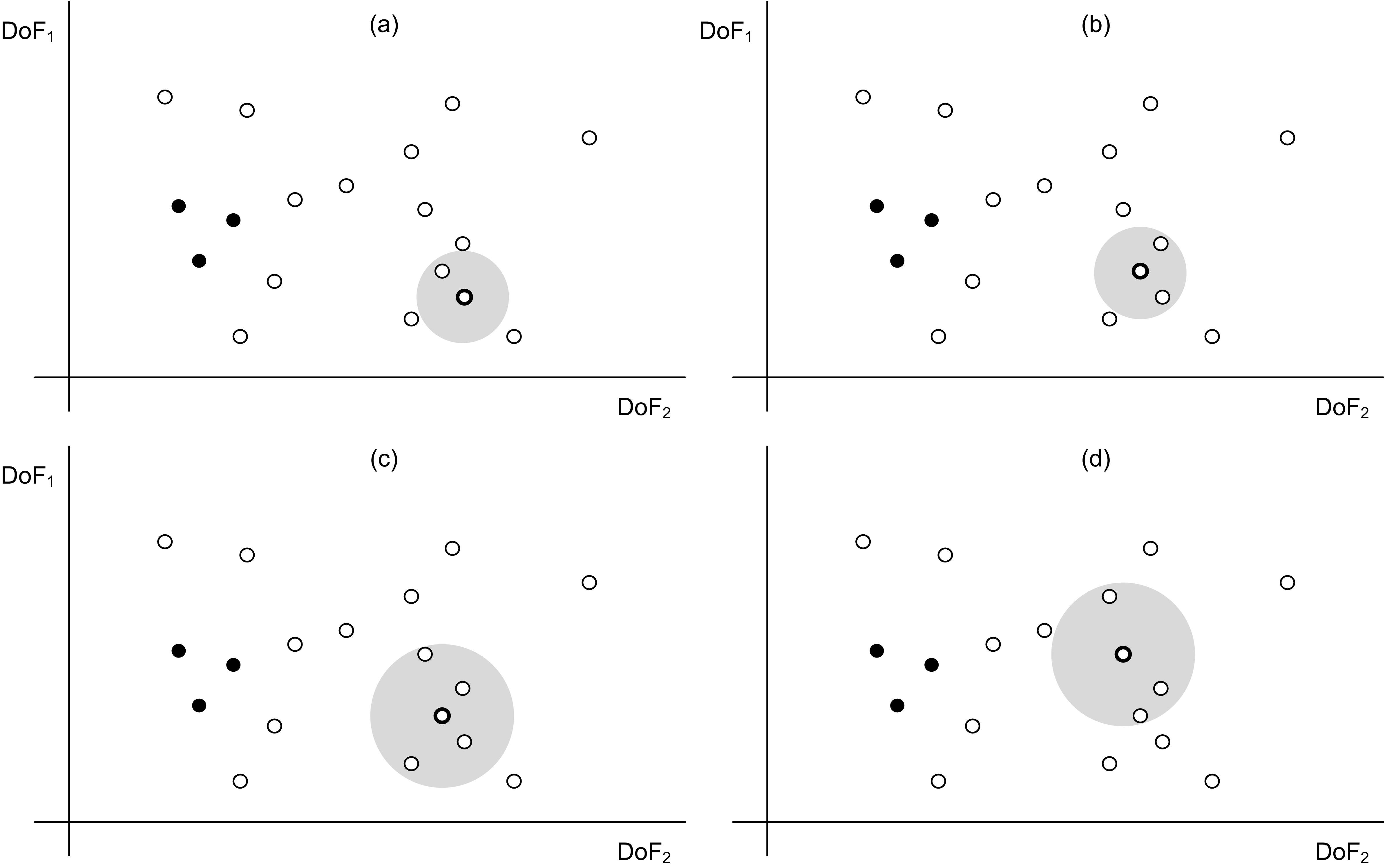}
	\caption{Run-Time Change of Actual Specification Point and Maximum Allowed Distance between the Actual Specification Point and a Proposed Specification Point. }
	\label{fig:spec-space-ASP}
	\end{center}
\end{figure}

Note that the maximum allowed distance between the actual point and a proposed point may change during the system execution.
%
%
Consider, for example, snapshots (a)--(d) of the Euclidean specification space with two DoF shown in Figure \ref{fig:spec-space-ASP}.
Black circles denote specification points representing specification instances that do not satisfy a set of protocol properties --- consequently, an agent is never empowered to propose that the actual specification point becomes one of those points (see Section \ref{sec:trans-prot}). 
White circles, on the other hand, denote specification points representing specification instances that satisfy the required protocol properties. The circle with the thick line denotes the actual specification point.
The grey area denotes the maximum allowed distance between the actual specification point and a proposed point --- this is expressed by $\mathit{threshold\_d}$.
Initially --- snapshot (a) --- only one specification point, $\mathit{sp}$, is within the grey area, that is, it is permitted, under certain circumstances, to move only to $\mathit{sp}$. Then --- snapshot (b) --- the actual specification point moves to $\mathit{sp}$. At this time, two specification points are within the grey area.  
Following this specification point change, the value of $\mathit{threshold\_d}$ increases --- snapshot (c) --- that is, the size of the grey area increases, offering more options for permitted specification point change. 
In some systems, designated agents, such as `institutional agents' \cite{bou08}, may increase, temporarily perhaps, the value of $\mathit{threshold\_d}$ in order to allow for a greater system specification change, possibly as a result of sensing a substantial change of environmental, social, or other conditions.
In other systems, changing the value of $\mathit{threshold\_d}$ may be realised in a manner similar to that used for changing the actual specification point.
Snapshot (d) shows that the actual specification point moves to a point that could not have been directly reached had the value of $\mathit{threshold\_d}$ not increased, assuming that agents abide by the rules governing specification point change. Note that, in general, the members of a system may adopt \emph{any} specification point that satisfies certain protocol properties --- in the present example the agents may adopt any point depicted as a white circle. Moving outside of the grey area, however, is forbidden and the agent that proposed such a specification point change may be subject to penalty.


The designers of a system may further constrain the process of run-time specification modification by permitting a proposal for specification point change only if the expected system utility associated with the proposed point is `acceptable' (in a sense to be specified below).
Consider the following formalisation:
\begin{align} 
\label{eq:per-propose-eu} 
&\begin{mysplit}
\mathit{\caused perPropose(Ag, NSP, PL) \IF}\\
       \qquad\mathit{powPropose(Ag, NSP, PL),}\\
       \qquad\mathit{actual\_sp(PL)\val ASP,}\\
	\qquad\mathit{distance( NSP, ASP, PL ) \leq  threshold\_d(PL), } \\
      \qquad\mathit{eu( NSP, PL ) \geq threshold\_eu(PL)}
\end{mysplit}\\
\label{eq:per-propose-eu2} 
&\begin{mysplit}
\mathit{\caused perPropose(Ag, NSP, PL) \IF}\\
       \qquad\mathit{powPropose(Ag, NSP, PL),}\\
       \qquad\mathit{actual\_sp(PL)\val ASP,}\\
	\qquad\mathit{distance( NSP, ASP, PL ) \leq  threshold\_d(PL), } \\
      \qquad\mathit{eu( NSP, PL ) > eu( ASP, PL )}
\end{mysplit}\\
\label{eq:per-propose-eu3} 
&\begin{mysplit}
\mathit{\default \neg perPropose(Ag, NSP, PL)}
\end{mysplit}
\end{align}	
The simple fluent constant $\mathit{eu(NSP, PL)}$ expresses the expected system utility associated with the specification point $\mathit{NSP}$ of level $\mathit{PL}$, that is, the system utility expected to be achieved under the specification instance corresponding to $\mathit{NSP}$. 
The utility of a system may be defined in various ways; in a resource-sharing protocol, for example, the system utility may be defined in terms of the average time for servicing a request for the floor. 
The simple fluent constant $\mathit{threshold\_eu(PL)}$ expresses the minimum allowed expected system utility in level $\mathit{PL}$.
The first three conditions of rules \eqref{eq:per-propose-eu} and \eqref{eq:per-propose-eu2} are the same as the conditions of rule \eqref{eq:per-propose-metric}. According to rules \eqref{eq:per-propose-eu}--\eqref{eq:per-propose-eu3}, an agent $\mathit{Ag}$ is permitted to propose that the specification point of level $\mathit{PL}$ becomes $\mathit{NSP}$, if and only if $\mathit{Ag}$ is empowered to make this proposal, $\mathit{NSP}$ is not too `far' from the actual point $\mathit{ASP}$, and 
\begin{itemize} 
 \item the expected system utility associated with $\mathit{NSP}$ exceeds a specified threshold, or
 \item the expected system utility associated with $\mathit{NSP}$ is greater than the expected system utility associated with $\mathit{ASP}$.
\end{itemize}

Rules \eqref{eq:per-propose-eu}--\eqref{eq:per-propose-eu3} are but one possible way to formalise the permission to propose a specification point change. We could have equally chosen to adopt, say, only rules \eqref{eq:per-propose-eu} and \eqref{eq:per-propose-eu3}, or rules \eqref{eq:per-propose-eu2} and \eqref{eq:per-propose-eu3}. 

Care should be taken when specifying the thresholds for distance and expected system utility. 
For example, it might be the case that, for certain values of these thresholds, it is never permitted to move from most specification points to any other point. 
%
%
\ccalc\ is very useful in selecting the appropriate values for these thresholds. 
We may use \ccalc\ to prove properties of a transition protocol under certain values of the distance and expected system utility thresholds, such as that, under certain circumstances it is permitted to move from some/all specification points to another point. In the next section we demonstrate the use of \ccalc\ for proving properties of an infrastructure for dynamic protocol specification, including a transition protocol specification.

\begin{figure}[t]
  \begin{center}
	\includegraphics[width=\textwidth]{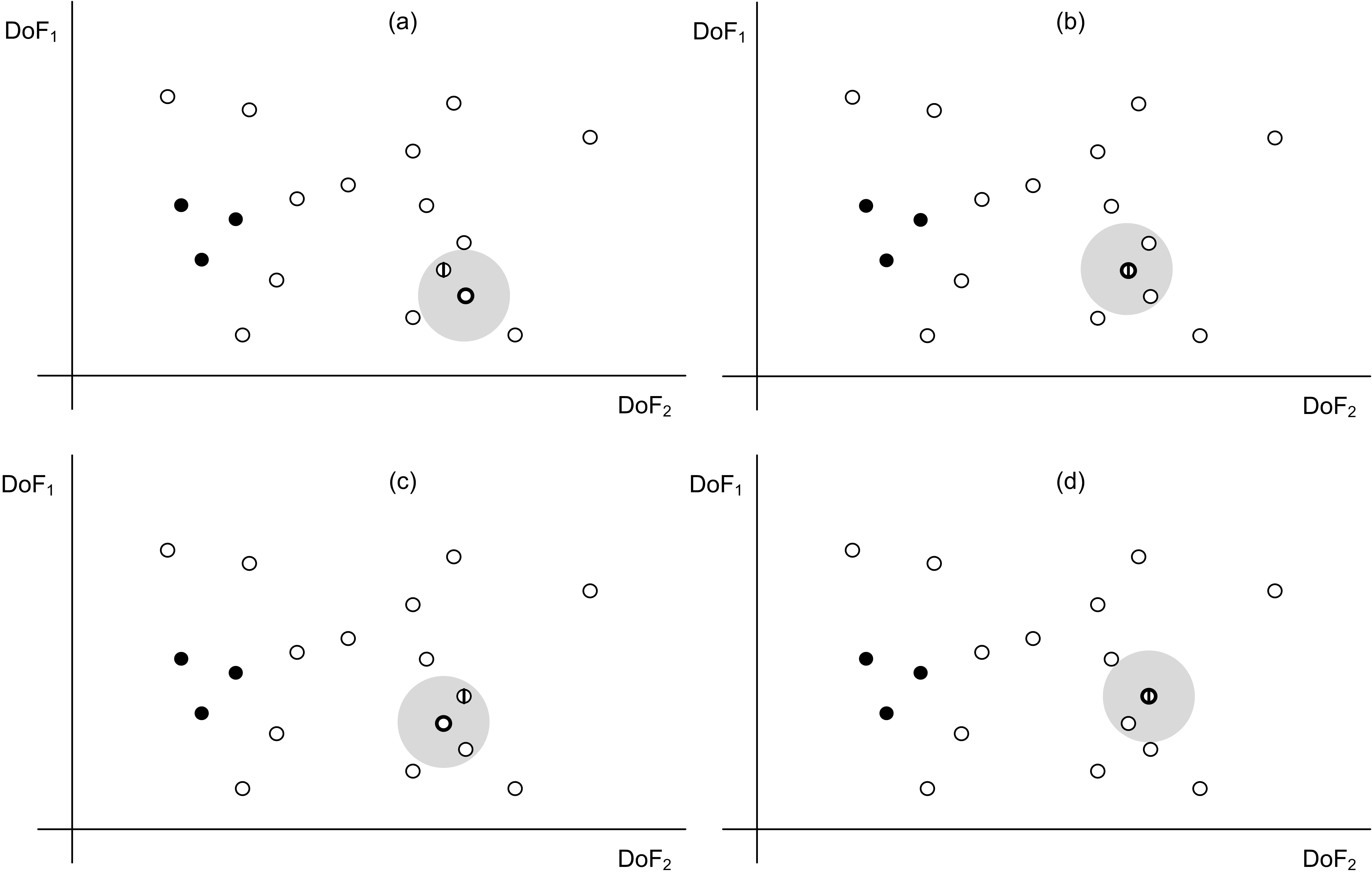}
	\caption{Run-Time Change of Desired and Actual Specification Points. }
	\label{fig:spec-space-DSP}
	\end{center}
\end{figure}

The system utility expected to be achieved under a specification instance depends on various conditions such as the size of the population of a system, the frequency of rule violation, and the available resources. Such conditions may fluctuate during a system's lifetime. Consequently, the system utility expected to be achieved under a specification instance may change over time.
Figure \ref{fig:spec-space-DSP} illustrates this issue; this figure shows four new snapshots of the Euclidean specification space presented in Figure \ref{fig:spec-space-ASP}. 
Recall that black circles denote specification points representing specification instances that do not satisfy a set of protocol properties, white circles denote specification points representing specification instances that do satisfy the required protocol properties, the circle with the thick line denotes the actual specification point, and the grey area denotes the maximum allowed distance between the actual specification point and a proposed point.
The circle with the vertical line denotes the `desired' specification point, that is, the specification point corresponding to the specification instance with the maximum expected system utility. To avoid clutter, we do not show the expected system utility of each specification point.
The snapshots of Figure \ref{fig:spec-space-ASP} show three specification point changes. First, the system members change the actual specification point in a way that it coincides with the desired point. Then, the desired specification point changes --- such a change could be the result of a change in environmental conditions, for example.
(In the present \cplus\ formalisation, we have written simple rules to compute the expected system utility associated with each specification point, and thus determine the desired point, under different environmental conditions. In other examples, institutional agents, or other types of agent may compute the expected system utility associated with each specification point, and therefore determine the desired point, under different conditions.)
Finally, the system members change once more the actual point in a way that it reaches the desired one.
Recall that the members of a system may adopt any specification point (that satisfies certain protocol properties). In other words, it may not necessarily be the case that the members of a system try to reach the desired specification point, or move to a point that increases the expected system utility.

Apart from the desired point, the threshold concerning expected system utility ($\mathit{threshold\_eu}$) may change over time, using institutional agents or some other means.


Similar to rules \eqref{eq:per-propose-eu}--\eqref{eq:per-propose-eu3}, we may express the permission to \emph{second} a proposal for specification point change, or the \emph{obligation to object} to such a proposal. Moreover, when the expected system utility associated with the actual specification point is below the specified threshold ($\mathit{threshold\_eu}$), we could impose an \emph{obligation to propose} a specification point change that, if accepted (the proposal), would change the actual point in a way that the associated expected system utility is increased.
Below is a way of formalising such an obligation:
\begin{equation} \label{eq:obl-propose} 
\begin{mysplit}
\mathit{\caused oblPropose(Ag, NSP, PL) \IF}\\
       \qquad\mathit{perPropose(Ag, NSP, PL),}\\
	\qquad\mathit{actual\_sp(PL)\val ASP,}\\
       \qquad\mathit{eu( ASP, PL ) < threshold\_eu(PL)}
\end{mysplit}
\end{equation}
The statically determined fluent constant $\mathit{oblPropose}$ expresses the obligation to propose a  specification point change.
According to the above rule, an agent $\mathit{Ag}$ is obliged to propose that the specification point of level $\mathit{PL}$ becomes $\mathit{NSP}$ if $\mathit{Ag}$ is permitted to propose the adoption of $\mathit{NSP}$ (see rules \eqref{eq:per-propose-eu}--\eqref{eq:per-propose-eu3} for an example definition of the permission to propose a specification point change), 
and the expected system utility associated with the actual specification point of level $\mathit{PL}$, $\mathit{ASP}$, is less than the $\mathit{threshold\_eu(PL)}$ value. 

Note that $\mathit{Ag}$'s obligation to propose a specification point change may be terminated, even if $\mathit{Ag}$ does not discharge it: a specification point with greater expected system utility may be adopted due to the proposal of some other agent. In this case the last condition of rule \eqref{eq:obl-propose} may cease to hold and thus $\mathit{Ag}$ will no longer be obliged to propose a specification point change.

\section{Proving Properties of the Dynamic Resource-Sharing Protocol}\label{sec:dynamic-properties}

Specifying a dynamic protocol in \cplus\ allows us to prove various properties of the specification. (Recall that in Section \ref{sec:cfcp-properties} we proved properties of a static resource-sharing protocol specification.) 
We may prove properties of the specification instances of level 0 and level $n$, $n{>}0$, and the transition protocols. For example, we may prove that the transition protocols and level $n$ protocols terminate within a fixed number of steps,
that a level $n$ protocol may be initiated at most a fixed number of times, and so on. 
Below we present a few example properties proven of the presented dynamic resource-sharing protocol, by means of \ccalc\ query computation.

Recall that the \cplus\ action description \drs, expressing the dynamic resource-sharing protocol, defines a three-level infrastructure; level 0 (resource-sharing) has three DoF, the specification of the best candidate for the floor, the permission to assign the floor, and the permission to request a resource manipulation.
Level 1 voting has a single DoF, the standing rules of the voting procedure, while level 2 voting has no DoF.
To conduct computational experiments, one has to make specific choices for a set of parameters. For a concrete illustration we will present here experiments in which, for example, the distance between two specification points is computed with the use of a weighted Manhattan metric. Arbitrary values were chosen for the threshold distance between the actual specification point and a proposed point, as well as the threshold expected system utility. 
Moreover, the only specification points representing inconsistent, in some sense, specification instances are the level 0 points of the form $\mathit{(rmt, *, any\_type)}$ (see Section \ref{sec:trans-prot}). These specification points may not be reached because, in this example, no agent is empowered to propose the adoption of a specification point representing an inconsistent specification instance.
%
%
Other choices concerning the experimental parameters could of course have been made, and the experiments repeated for those.

\begin{Property} 
\label{prop:propd1}
According to the specification instance corresponding to specification point $\mathit{sp_2}$ of level 0, there is no protocol state in which a subject is permitted to request a resource manipulation type different from that stated when applying for access to the resource.
\end{Property}

We instruct \ccalc\ to compute all states $s$ of \drs\ such that 
\begin{equation*} 
	\mathit{s\ \models\ actual\_sp(0)\val sp_2\ \wedge\ perRequestMpt(S, FCS, M) }
\end{equation*}
In other words, we are interested in computing all states in which the actual specification point of level 0 (resource-sharing) is $\mathit{sp_2}$ (see expression \eqref{eq:bc-configB}), and a subject is permitted to request from the Floor Control Server (FCS) to manipulate the resource. 

In all solutions computed by \ccalc\ we have that 
\begin{equation*} 
	\mathit{s\ \models\ perRequestMpt( S, FCS, requested(S) ) }
\end{equation*}
Recall that the value of a simple fluent constant $\mathit{requested(S)}$ is the resource manipulation type (say, storing files of a particular type on the shared storage device) stated by subject $S$ when it applied for access to the resource. According to the solutions produced by \ccalc, a subject $S$ is permitted to request from FCS only the type of resource manipulation expressed when $S$ applied for access to the resource. 
This is due to $\mathit{dof(per\_mpt, sp_2)\val expressed\_type}$ of expression \eqref{eq:bc-configB} --- recall that $\mathit{per\_mpt}$ represents the DoF concerning the permission to request a resource manipulation type --- and rule \eqref{eq:permpt-req}.

\vspace{9pt} 

\begin{Property} 
\label{prop:propd2}
There is no protocol state in which an agent is forbidden and obliged to propose a specification point change.
\end{Property}

We instruct \ccalc\ to compute all states $s$ of \drs\ such that
\begin{equation*}
\mathit{s \models oblPropose(Ag, NSP, PL)}
\end{equation*}

For all states $s$ computed by \ccalc\ we obtain
\begin{equation*}
\mathit{s \models perPropose(Ag, NSP, PL) }
\end{equation*}
that is, there is no state in which an agent is obliged and forbidden to propose a specification point change.
%
In the presented computational experiments, rules \eqref{eq:per-propose-eu}--\eqref{eq:per-propose-eu3} express the conditions in which an agent is permitted to propose a specification point change, while rule \eqref{eq:obl-propose} expresses the conditions in which an agent is obliged to propose a specification point change. 

\vspace{9pt} 

\begin{Property} 
\label{prop:propd3}
Exercising the power to declare the outcome of the voting procedure of level $n$ carried always changes the specification point of level $n{-}1$.
\end{Property}

We instruct \ccalc\ to find all states $s'$ such that 
\begin{itemize}
\item $\mathit{(s, \e, s')}$ is a transition of \drs, 
\item $\mathit{s\ \models\ powDeclare(VC, NSP, carried, PL)\ \wedge\ actual\_sp(PL{-}1)\val ASP}$, and 
\item $\mathit{\e\ \models\ declare(VC, NSP, carried, PL)}$.
\end{itemize}

For every state $s'$ computed by \ccalc\ we obtain
\begin{equation*} 
\mathit{s'\ \models\ actual\_sp(PL{-}1)\val NSP}
\end{equation*}
which denotes that the specification point of level $\mathit{PL}$ has changed (in all solutions $\mathit{ASP\neq NSP}$). Rule \eqref{eq:declare-replace} expresses the effects of exercising the power to declare the outcome of a voting procedure carried.

In some solutions to the above query, the subjects are obliged in the resulting state $s'$ to propose a specification point change in level $PL{-}1$. This is due to the fact that, in these solutions, the expected system utility associated with the new specification point, $\mathit{NSP}$, is below the specified threshold (that is, $\mathit{eu(NSP, PL{-}1) < threshold\_eu(PL{-}1)}$).

\section{Animating the Dynamic Resource-Sharing Protocol}\label{sec:animation}

Apart from proving properties of a dynamic protocol specification, \ccalc's query computation --- in particular the computation of `prediction' (temporal projection) queries, such as the queries used to prove Properties \ref{prop:prop2} and \ref{prop:propd3} --- allows us to calculate, at run-time, the agents' powers, permissions, and obligations.
Such information may be publicised to the members of a system, and may be provided by a central server or in various distributed configurations. (Further discussion of these architectural issues is outside the scope of this paper.)
In this section we present an example execution (run) of the dynamic resource-sharing protocol, and the results obtained by \ccalc's prediction query computation. 

The narrative of events of the presented run is displayed in Table \ref{tbl:cfcp-exec}. 
The events of transition protocols, $\mathit{propose}$, $\mathit{second}$, $\mathit{object}$, level 1 and level 2 protocols, $\mathit{vote}$ and $\mathit{declare}$, are indented. 
The last argument of a level 1 or 2 event indicates the protocol level in which the event took place. To save space, we group the votes that are in favour of (respectively, against) a motion.
In the initial state of the presented run:

\begin{itemize}
\item The actual specification point of level 0 is $\mathit{(fcfs, 3, any\_type)}$, that is, the best candidate for the floor is determined on a first-come, first-served basis, the maximum number of permitted consecutive allocations of the floor is 3, while the holder is permitted to request any type of resource manipulation (see rule \eqref{eq:permpt-any}).

\item The actual specification point of level 1 is $\mathit{(3\_4m)}$, that is, a 75\% majority is required (recall that level 1 has a single DoF).

\end{itemize} 
Level 2 voting does not have a DoF. In these experiments level 2 voting requires a 75\% majority. 
Details about the choices we made concerning the remaining experimental parameters were given in the previous section.

\begin{table}[t]
\caption{Run of Dynamic Resource-Sharing Protocol.}\label{tbl:cfcp-exec}
\renewcommand{\arraystretch}{1}
\setlength\tabcolsep{3pt}
\begin{tabular}{cl}
\hline\noalign{\smallskip}
\multicolumn{1}{c}{\textbf{Time}} & \multicolumn{1}{c}{\textbf{Action}} \\
\noalign{\smallskip}
\hline
\noalign{\smallskip}
0 & $\mathit{request\_floor(sub_1,\ c,\ app_A)}$   \\   
5 & $\mathit{request\_floor(sub_2,\ c,\ app_A)}$   \\   
6 & $\mathit{request\_floor(sub_3,\ c,\ app_A)}$   \\   
8 & $\mathit{request\_floor(sub_5,\ c,\ app_A)}$   \\   
14 & $\mathit{\quad propose(sub_3,\ sp_{26},\ 1)}$   \\
16 & $\mathit{\quad object(sub_1,\ sp_{26},\ 1)}$   \\
17 & $\mathit{\quad second(sub_5,\ sp_{26},\ 1)}$   \\
 & $\quad$ transition protocol argumentation  \\
28 & $\mathit{\qquad\qquad vote([sub_2, sub_3, sub_4, sub_5, sub_6],\ for,\ 2)}$   \\
30 & $\mathit{\qquad\qquad vote(sub_1,\ against,\ 2)}$   \\
31 & $\mathit{\qquad\qquad declare(c,\ sp_{26},\ carried,\ 2)}$   \\

35 & $\mathit{\quad propose(sub_5,\ sp_3,\ 0)}$  \\
36 & $\mathit{\quad second(sub_3,\ sp_3,\ 0)}$  \\
39 & $\mathit{\quad object(sub_2,\ sp_3,\ 0)}$  \\
 & $\quad$  transition protocol argumentation   \\
51 & $\mathit{\qquad\quad vote([sub_3, sub_4, sub_6],\ for,\ 1)}$   \\
53 & $\mathit{\qquad\quad vote([sub_1, sub_2],\ against,\ 1)}$   \\
54 & $\mathit{\qquad\quad declare(c,\ sp_{3},\ carried,\ 1)}$  \\
58 & $\mathit{assign\_floor(c,\ sub_3)}$  \\

\hline
\end{tabular}
\end{table}


The present example includes 7 agents, a chair $c$ and 6 subjects \sba--\sbf. In the beginning of the run-time activities \sba, \sbb, \sbc\ and \sbe\ exercise their power to request the floor, all of them requiring to run applications of type $A$ on the shared processor (see the third argument of $\mathit{request\_floor}$). \sbc\ and \sbe\ aim to change the specification point of level 0 in a way that the value of the best candidate DoF becomes $\mathit{random}$, that is, the best candidate for the floor is chosen randomly from the list of subjects having pending floor requests. In this way \sbc\ and \sbe\ may acquire the floor faster. Before attempting to change the specification point of level 0, \sbc\ and \sbe\ attempt to change the specification point of level 1 in a way that level 1 voting requires simple majority as opposed to 75\% majority. Therefore, fewer votes will be required in level 1 when \sbc\ and \sbe\ propose to change the specification point of level 0. The proposal for changing the specification point of level 1 takes place at time-point 14 --- specification point $\mathit{sp_{26}}$ expresses that the standing rules require simple majority. At that time \sbc\ is empowered to make the proposal (see rule \eqref{eq:propose-replace}).
Furthermore, \sbc\ is permitted to exercise its power (see rules \eqref{eq:per-propose-eu}--\eqref{eq:per-propose-eu3}) because: (i) the distance, in level 1, between the actual specification point (75\% majority) and the proposed point $\mathit{sp_{26}}$ (simple majority) is less than the chosen threshold ($\mathit{threshold\_d(1)}$), and (ii) the expected system utility associated with $\mathit{sp_{26}}$ is greater than the corresponding threshold ($\mathit{threshold\_eu(1)}$).
\sbc's proposal is followed by an objection, a secondment, and an argumentation. Then level 2 voting commences; the motion is the adoption of $\mathit{sp_{26}}$ in level 1. \sbb--\sbf\ vote for the motion while \sba\ votes against it. At time-point 31 the motion of level 2 is declared carried (recall that level 2 requires 75\% majority) and thus the specification point of level 1 becomes $\mathit{sp_{26}}$ (see rule \eqref{eq:declare-replace}), meaning that the standing rules of level 1 change to simple majority. 

\sbe\ proposes at time-point 35 the adoption of specification point \linebreak $\mathit{sp_3 \val (random, 3, any\_type)}$ in level 0. \sbe\ is empowered to make the proposal at that time. However, \sbe\ is not permitted to exercise its power because the expected system utility of $\mathit{sp_3}$ is less than $\mathit{threshold\_eu(0)}$, and less than the expected system utility associated with the actual point $\mathit{(fcfs, 3, any\_type)}$ of level 0. Consequently, \sbe\ is sanctioned for performing a forbidden action and thus cannot participate in level 1 to vote (see rule \eqref{eq:meta-role-ass}). \sbc, \sbd\ and \sbf\ vote for the motion of level 1 --- the adoption of $\mathit{sp_3}$ in level 0 --- while \sba\ and \sbb\ vote against it. At time-point 54 the motion of level 1 is declared carried (recall that level 1 now requires a simple majority) and thus the specification point of level 0 becomes $\mathit{sp_3}$, meaning that the best candidate for the floor is chosen randomly from the list of subjects having pending floor requests. At time 58 $c$ exercises its power (see rule \eqref{eq:pow-assign}) to assign the floor to \sbc.

\section{Summary, Related and Further Work}\label{sec:discussion}

We presented an infrastructure for dynamic specifications for open MAS, that is, specifications that are developed at design-time but may be modified at run-time by the members of a system. Any protocol for open MAS may be in level 0 of our infrastructure, whereas any protocol for decision-making over specification change may be in level $n$, $n{>}0$. The level 0/level $n$/transition protocols can be as complex/simple as required by the application in question.

We employed \cplus, an action language with explicit transition system semantics to formalise dynamic specifications. Moreover, we employed \ccalc, an automated reasoning tool for proving properties of the specifications and assimilating narratives of events. 
On the one hand, therefore, we may provide design-time services, proving properties, such as termination, of the various protocols of our infrastructure, and on the other hand, we may offer run-time services, calculating the system state current at each time, including the powers, permissions and obligations of the agents.



Chopra and Singh \cite{chopra06} have presented a way of adapting `commitment protocols' \cite{singh98b, singh99a, venkatraman99, singh00} according to context, or the preferences of agents in a given context. They formalise protocols and `transformers', that is, additions/enhancements to an existing protocol specification that handle some aspect of context or preference. Depending on the context or preference, a protocol specification is complemented, at \emph{design-time}, by the appropriate transformer thus resulting in a new specification. Unlike Chopra and Singh, we are concerned here with the \emph{run-time} adaptation of a protocol specification and, therefore, we developed an infrastructure --- meta protocols, transition protocols --- to achieve that.

Several approaches have been proposed in the literature for run-time specification change of norm-governed MAS. Serban and Minsky \cite{minsky09}, for example, have presented a framework for law change in the context of `Law-Governed Interaction' (LGI) \cite{minsky91b, minsky00, minsky03, minsky07global, minsky08bac}. LGI is an abstract regulatory mechanism that satisfies the following principles: statefulness, that is, the regulatory mechanism is sensitive to the history of interaction between the regulated components, decentralisation, for scalability, and generality, that is, LGI is not biased to a particular type of law. LGI is an abstraction of a software mechanism called Moses \cite{minsky05moses, minsky04open} which can be used to regulate distributed systems. Moses employs regimentation devices that monitor the behaviour of agents, block the performance of forbidden actions and enforce compliance with obligations. 

It has been argued \cite{jones93} that regimentation is rarely desirable (it results in a rigid system that may discourage agents from entering it \cite{prakken98}), and not always practical. In any case, violations may still occur even when regimenting a MAS (consider, for instance, a faulty regimentation device). For all of these reasons, we have to allow for non-compliance and sanctioning and not rely exclusively on regimentation mechanisms.

Serban and Minsky were concerned in \cite{minsky09} with architectural issues concerning law change. They presented a framework with which a law change is propagated to the distributed regimentation devices, taking into consideration the possibility that during a `convergence period' various regimentation devices operate under different versions of a law, due to the difficulties of achieving synchronised, atomic law update in distributed systems. 

There are several other approaches in the literature concerned with architectural issues of run-time specification change --- \cite{ponder01, chandha04, godic05} are but a few examples. 
These issues --- various distributed configurations for computing the normative relations current at each time --- are beyond the scope of this paper, and will be considered in future work.


Bou and colleagues \cite{bou08, bou07} have presented a mechanism for the run-time modification of the norms of an `electronic institution' \cite{esteva00, esteva01, esteva01a, esteva02a, rodriguez02a, VESR04, GNR05}. These researchers have proposed a `normative transition function' that maps a set of norms (and goals) into a new set of norms: changing a norm requires changing its parameters, or its effect, or both. The `institutional agents', representing the institution, are observing the members' interactions in order to learn the normative transition function, so that they (the institutional agents) will directly enact the norms enabling the achievement of the `institutional goals' in a given scenario. Unlike Bou and colleagues, we do not necessarily rely on designated agents to modify norms. We presented an infrastructure with which any agent may (attempt to) adapt the system specification. This does not exclude the possibility, however, that, in some applications, designated agents are given, under certain circumstances, the institutional power to directly modify the system specification.

Identifying \emph{when} (to propose) to change a system specification during the system execution, as done in the work of Bou and colleagues with respect to the `institutional goals', is a fundamental requirement for adaptive MAS. Addressing this requirement, however, is out of the scope of this paper.


Boella and colleagues \cite{boella09aamas} have developed a formal framework for representing norm change --- see \cite{broersen09oamas} for a recent survey on formal models of norm change. The framework of norm change presented in \cite{boella09aamas} is produced by replacing the propositional formulas of the Alchourr\'{o}n, G\"{a}rdenfors, Makinson (AGM) framework of theory change \cite{alchourron85} with \emph{pairs} of propositional formulas --- the latter representing norms --- and adopting several principles from input/output logic \cite{makinson00}. The resulting framework includes a set of postulates defining norm change operations, such as norm contraction. 

Governatori and colleagues \cite{governatori07icail, governatori08DEON, governatori08NORMAS} have presented variants of a Temporal Defeasible Logic \cite{governatori05JURIX, governatori06PRICAI} to reason about different aspects of norm change. These researchers have represented meta norms describing norm modifications by referring to a variety of possible time-lines through which the elements of a norm-governed system, and the conclusions that follow from them, can or cannot persist over time. Governatori at al.~have formalised norm change operations according to which norms are removed with all their effects, as well as operations according to which norms are removed  but all or some of their effects propagate if obtained before the modification. 

\cplus, along with proposed extensions \cite{craven05} of this language, allows for the formalisation of relatively complex norm change operations.  
We expect that the expressiveness of \cplus\ is adequate for representing norm change operations for a wide range of software MAS.
The infrastructure presented in this paper, however, may be formalised using other languages.
We chose \cplus\ because it enables the formal representation of the (direct and indirect) effects of actions and default persistence (inertia) of facts, has a transition system semantics and thus there is a link to a wide range of other formalisms and tools based on transition systems (later in this section we describe a way to  exploit this link by combining \cplus\ with standard model checkers), and has direct routes to implementation.

Rubino and Sartor \cite{rubino08} have presented a taxonomy of `source norms', that is, norms empowering the members of a system to modify a set of other norms.
The so-called `agreement-based source norms' may be viewed as the norms of a meta protocol in our infrastructure, in the sense that they allow for norm change based on the deliberation of a group of agents.
Rubino and Sartor employ the logic of the PRATOR system  for defeasible argumentation \cite{prakken97} to express source norms.
The formalisation of the ideas presented in \cite{rubino08}, however, is restricted; consequently, this line of work cannot be used yet to support dynamic MAS.


Run-time specification change has long been studied in the field of argumentation.  Loui \cite{loui92}, for example, identified the need to allow for the run-time modification of argumentation protocol rules by means of meta argumentation. Vreeswijk \cite{vreeswijk00} also investigated forms of meta argumentation. The starting point for this work was two basic observations. Firstly, that there are different protocols appropriate for different contexts (for example, quick and shallow reasoning when time is a constraint; restricted number of counter-arguments when there are many rules and cases; etc). Secondly, that `points of order', by which a participant may steer the protocol to a desired direction, are standard practice in dispute resolution meetings. Vreeswijk then defined a formal protocol for disputes in which points of order can be raised to allow (partial) protocol changes to be debated. A successful `defence' meant that the parties in the dispute agreed to adopt a change in the protocol, and the rules of dispute were correspondingly changed.


As already mentioned, our work is motivated by Brewka's dynamic argument systems \cite{brewka01}. Like Vreeswijk's work, these are argument systems in which the protagonists of a disputation may start a meta level debate, that is, the rules of order become the current point of discussion, with the intention of altering these rules. Unlike Vreeswijk's work, there may be more than one meta argumentation level.

A key difference between our work and Brewka's approach, and more generally, a key difference between our work and related research, including all approaches discussed in this section, is that we formalise the transition protocol leading from an object protocol to a meta protocol.
More precisely, we distinguish between successful and unsuccessful attempts to initiate a meta protocol (exercising the institutional power to propose/second/object to a specification change vs proposing/seconding/objecting to a specification change without the necessary power), evaluate proposals for specification change by modelling a specification as a metric space, and by taking into consideration the effects of accepting a proposal on system utility, constrain the enactment of proposals that do not meet the evaluation criteria, and formalise procedures for role-assignment in a meta level.

In this section we focused on the facilities offered by related approaches with respect to system specification \emph{change}.
A comparison of our work with commitment protocols (including \cite{colombetti00, vigano07, fornara07, fornara08, fornara08IGI}), LGI, electronic institutions, Brewka's argument systems, as well as other approaches for specifying norm-governed systems, from the viewpoint of \emph{static} specifications, may be found in \cite{artikisIGPL, artikis06AIJ}.

We have been concerned with a particular aspect of `organised adaptation' \cite{oamas08}: the run-time modification of the `rules of the game' of norm-governed systems. Clearly there are other aspects of (organised) adaptation such as the run-time alteration of the (trading and other) relationships between agents, the assignment of roles to agents, and the goals of a system. \cite{shoham97a, Tesauro:2004p5819, hubner04, excelente04, martin06, hoogendoorn07, hoogendoorn07ijcai, Kota:2009p5875, deloach08, deloach08hbook, costa08book} are but a few examples of studies of adaptive systems.


We employed \ccalc\ for the provision of run-time services --- for instance, to compute the system state current at each time. \ccalc, however, can become inefficient when considering action descriptions defining large transition systems. (\ccalc\ is not the only means by which \cplus\ action descriptions can be executed. In \cite{lifschitz01} it is shown how \cplus\ action descriptions can be translated into the formalism of (extended) logic programs.)
We have also used versions of the Event Calculus (EC) \cite{kowalski86} to specify and execute dynamic specifications \cite{artikis09aamas}. EC is a simple and flexible formalism that is efficiently implemented for narrative assimilation. Our Prolog EC implementation, however, does not offer facilities for proving properties of a specification or planning. More importantly, we also lose the explicit transition system semantics which we see as a very important advantage of the \cplus\ formulation. A discussion comparing the use of EC and \cplus\ for developing executable MAS specifications can be found in \cite{artikisToCL}.

A direction for further work is to employ a single formalism for efficient execution (narrative assimilation and proving properties) of (dynamic) MAS specifications. Some first steps are reported by Craven \cite{craven:thesis} who investigates methods for efficient EC-like query evaluation for (a subset of) the \cplus\ language, and for integrating action descriptions in this language with standard model checking systems (specifically NuSMV \cite{NuSMV2}). Norm-governed system properties expressed in temporal logics such as Computation Tree Logic can then be verified by means of standard model checking techniques on a transition system defined using the \cplus\ language.

\section*{Acknowledgements}

This paper is a significantly updated and extended version of \cite{artikis09aamas, artikis09icail}. We would like to thank the reviewers and participants of the Eighth International Conference on Autonomous Agents and Multi-Agent Systems and the Twelfth International Conference on Artificial Intelligence \& Law, who gave us useful feedback. We would also like to thank Lloyd Kamara for his comments on various drafts of the paper.


\bibliographystyle{plain}

\end{document}